\documentclass[10pt,aps,prd,nofootinbib,superscriptaddress,twocolumn,preprintnumbers,balancelastpage]{revtex4}

\usepackage{amssymb,amsmath,latexsym,graphics, graphicx,epsfig,multirow,comment,appendix,feyn,slashed,color,afterpage,multirow,makecell} 

\usepackage[colorlinks=true
,urlcolor=blue
,anchorcolor=blue
,citecolor=blue
,filecolor=blue
,linkcolor=blue
,menucolor=blue
,linktocpage=true
,pdfproducer=medialab
,pdfa=true
]{hyperref}

\usepackage{tabularx}

\newcommand {\be} {\begin {equation}}
\newcommand {\ee} {\end {equation}}

\newcommand {\bes} {\begin {equation*}}
\newcommand {\ees} {\end {equation*}}

\newcommand{\es}[2] {\begin{equation} \label{#1} \begin{split} #2 \end{split} \end{equation}}

\def\CO{{\cal O}}

\newcommand{\beq}{\begin{equation}}
\newcommand{\eeq}{\end{equation}}

\begin{document}

\title{
Gamma-ray Constraints on Decaying Dark Matter and Implications for IceCube  
}
\preprint{
MIT-CTP/4863}

\author{Timothy Cohen}
\affiliation{Institute of Theoretical Science, University of Oregon, Eugene, OR 97403}

\author{Kohta Murase}
\affiliation{Center for Particle and Gravitational Astrophysics; Department of Physics;\\
 Department of Astronomy and Astrophysics, \\
 The Pennsylvania State University, University Park, Pennsylvania 16802}
\affiliation{Yukawa Institute for Theoretical Physics, Kyoto University, Kyoto 606-8502, Japan}

\author{Nicholas L. Rodd}
\affiliation{Center for Theoretical Physics, Massachusetts Institute of Technology, Cambridge, MA 02139}

\author{Benjamin R. Safdi}
\affiliation{Center for Theoretical Physics, Massachusetts Institute of Technology, Cambridge, MA 02139}

\author{Yotam Soreq}
\affiliation{Center for Theoretical Physics, Massachusetts Institute of Technology, Cambridge, MA 02139}

\begin{abstract}
\begin{center}
{\bf Abstract}
\vspace{-7pt}
\end{center}
Utilizing the {\it Fermi} measurement of the gamma-ray spectrum toward the Inner Galaxy, we derive some of the strongest constraints to date on the dark matter~(DM) lifetime in the mass range from hundreds of MeV to above an EeV.  Our profile-likelihood based analysis relies on 413 weeks of {\it Fermi} Pass 8 data from 200 MeV to 2 TeV, along with up-to-date models for diffuse gamma-ray emission within the Milky Way.  We model Galactic and extragalactic DM decay and include contributions to the DM-induced gamma-ray flux resulting from both primary emission and inverse-Compton scattering of primary electrons and positrons.  For the extragalactic flux, we also calculate the spectrum associated with cascades of high-energy gamma-rays scattering off of the cosmic background radiation. We argue that a decaying DM interpretation for the 10~TeV-1~PeV neutrino flux observed by IceCube is disfavored by our constraints. Our results also challenge a decaying DM explanation of the AMS-02 positron flux. We interpret the results in terms of individual final states and in the context of simplified scenarios such as a hidden-sector glueball model.
\end{abstract}
\maketitle

  \begin{figure}[t]
	\leavevmode
	\vspace{-.30cm}
	\begin{center}
        \includegraphics[width = 0.95 \columnwidth]{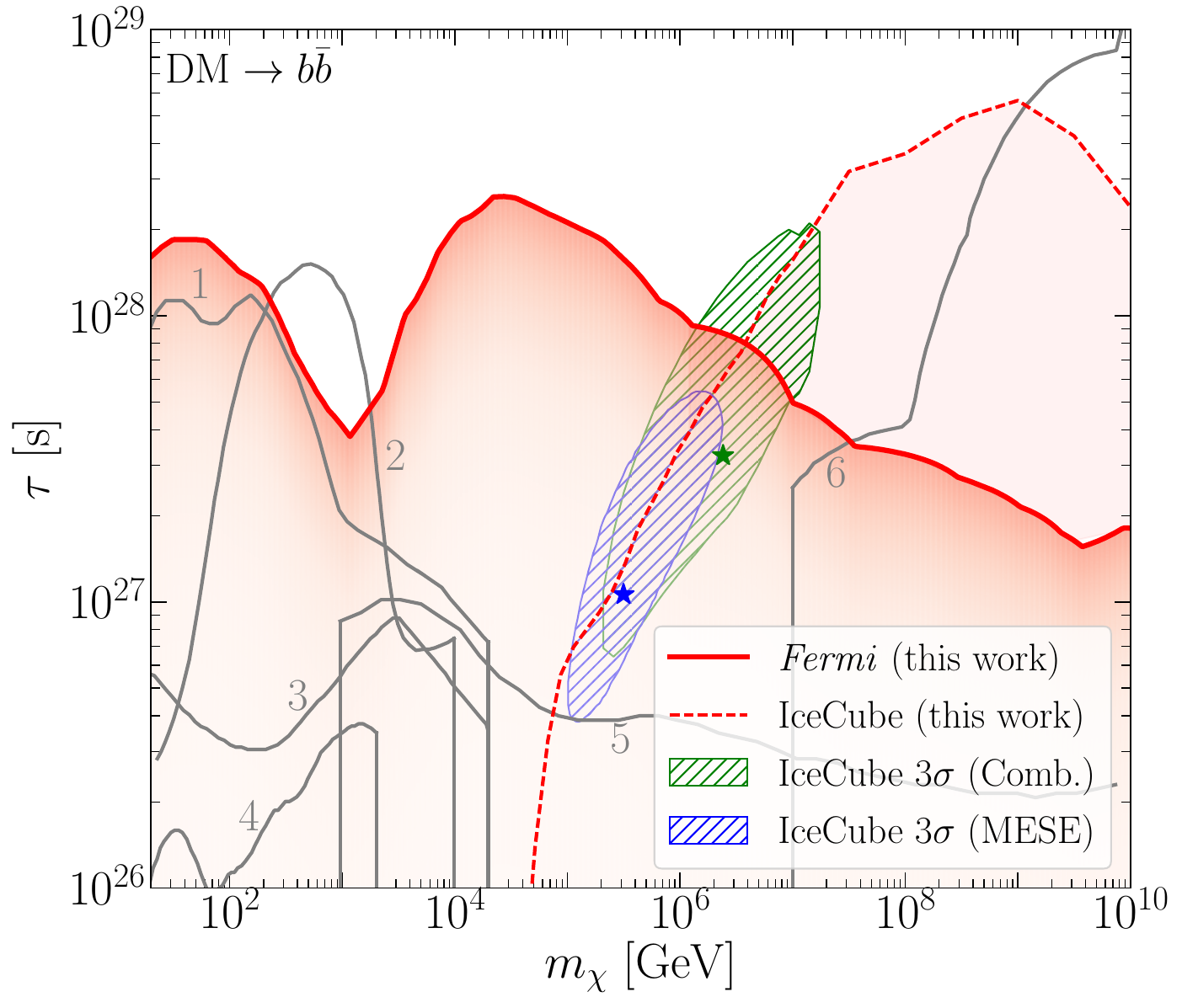}
	\end{center}
	\vspace{-.50cm}
	\caption{Limits 
	on DM decays to $b\, \bar{b}$, as compared to previously computed limits using data from {\it Fermi} (2,3,5), AMS-02 (1,4), and PAO/KASCADE/CASA-MIA  (6). 
	The hashed green (blue) region suggests parameter space where DM decay may provide a $\sim$$3 \sigma$ 
	improvement to the description of the combined maximum likelihood (MESE) IceCube neutrino flux.  The best-fit points, marked as stars, are in strong tension with our gamma-ray results.  The red dotted line provides a limit if we assume a combination of DM decay and astrophysical sources are responsible for the spectrum. 
	}
	\vspace{-0.15in}
	\label{Fig: bbResult}
\end{figure} 

A primary goal of the particle physics program is to discover the connection between dark matter~(DM) and the Standard Model~(SM).  While the DM is known to be stable over cosmological timescales, rare DM decays may give rise to observable signals in the spectrum of high-energy cosmic rays.  Such decays would be induced through operators involving both the dark sector and the SM.  In this work, we derive some of the strongest constraints to date on decaying DM for masses from $\sim$400 MeV to $\sim$$10^{7}$\,GeV by performing a dedicated analysis of \emph{Fermi} gamma-ray data from 200\,MeV to 2\,TeV.

The solid red line in Fig.~\ref{Fig: bbResult} gives an example of our constraint on the DM~($\chi$) lifetime, $\tau$, as a function of its mass, $m_\chi$, assuming the DM decays exclusively to a pair of bottom quarks. Our analysis includes three contributions to the photon spectrum:  (1) prompt emission, (2) gamma-rays that are up-scattered by primary electrons/positrons through inverse Compton (IC) within the Galaxy, and (3) extragalactic contributions. 

In addition to deriving some of the strongest limits on the DM lifetime across many DM decay channels, our results provide the first dedicated constraints on DM using the latest \emph{Fermi} data for $m_\chi \gtrsim 10$\,TeV.  To emphasize this point, we provide a comparison with other limits in Fig.~\ref{Fig: bbResult}.   The dashed red curve indicates our new estimate of the limits set by high-energy neutrino observations at the IceCube experiment~\cite{Aartsen:2013bka,Aartsen:2013jdh,Aartsen:2015knd,Aartsen:2015rwa}. Our IceCube constraint dominates in the range from $\sim$$10^7$ to $10^9$\,GeV.  

Constraints from previous studies are plotted as solid grey lines labeled from 1-6.  Curve 6 shows that for masses above $\sim$$10^9$~GeV, limits from null observations of ultra-high-energy gamma-rays at air shower experiments~\cite{Kalashev:2016cre}, such as the Pierre Auger Observatory (PAO)~\cite{Aab:2015bza}, KASCADE~\cite{Kang:2015gpa}, and CASA-MIA~\cite{Chantell:1997gs}, surpass our IceCube limits. Curves 2, 5, and 3 are from previous analyses of the extragalactic~\cite{Ando:2015qda,Murase:2012xs} and Galactic~\cite{Ackermann:2012rg} {\it Fermi} gamma-ray flux  (for related work see~\cite{Hutsi:2010ai,Cirelli:2012ut,Kalashev:2016xmy}).  Our results are less sensitive to astrophysical modeling than~\cite{Ando:2015qda}, which makes assumptions about the classes of sources and their spectra that contribute to the unresolved component of the extragalactic gamma-ray background.  We improve and extend beyond~\cite{Ackermann:2012rg,Murase:2012xs} in a number of ways: by including state-of-the-art modeling for cosmic-ray-induced gamma-ray emission in the Milky Way, a larger and cleaner data set, and a novel analysis technique that allows us to search for a combination of Galactic and extragalactic flux arising from DM decay. The limits labeled 1 and 4 in Fig.~\ref{Fig: bbResult} are from the AMS-02 antiproton~\cite{ams02pos,Giesen:2015ufa} and positron~\cite{Aguilar:2013qda,Ibarra:2013zia} measurements, respectively; these constraints are subject to considerable astrophysical uncertainties, due to the propagation of charged cosmic rays from their source to Earth.

An additional motivation for this work is the measurement of the so far unexplained high-energy neutrinos observed by the IceCube experiment~\cite{Aartsen:2013bka,Aartsen:2013jdh,Aartsen:2015knd,Aartsen:2015rwa}. If the DM has both a mass \mbox{$m_\chi \sim$} PeV and a long lifetime \mbox{$\tau \sim$ $10^{28}$} seconds, its decays could contribute to the upper end of the IceCube spectrum. These DM candidates would produce correlated cosmic-ray signals, yielding a broad spectrum of gamma rays with energies extending well into \emph{Fermi}'s energy range. Taking this correlation between neutrino and photon spectra into account enables us to constrain the DM interpretation of these neutrinos using the \emph{Fermi} data.  

Figure~\ref{Fig: bbResult} illustrates regions of parameter space where we fit a decaying DM spectrum to the high-energy neutrino flux at IceCube in hashed green. The corresponding region for the analysis of Ref.~\cite{Chianese:2016kpu} using lower-energy neutrinos is shown in blue. Clearly, much of the parameter space relevant for IceCube is disfavored by the gamma-ray limits; the best fit points (indicated by stars) are in strong tension with the {\it Fermi} observations.  We conclude that models where decaying DM could account for the entire astrophysical neutrino flux observed by IceCube are disfavored.  Furthermore, models where the neutrino flux results from a mix of decaying DM and astrophysical sources are strongly constrained.  

The rest of this Letter is organized as follows.  First, we discuss the various contributions to the gamma-ray flux resulting from DM decay.  Then, we give an overview of the data set and analysis techniques used in this work.  Next, we provide context for these limits by interpreting them as constraints on a concrete model (glueball DM), before concluding.

\setcounter{section}{1}
\vspace{-8pt}
\section{The Gamma-ray flux}
\vspace{-10pt}
\label{sec:CalcGammaFlux}
Decaying DM contributes both a Galactic and extragalactic flux.  The Galactic contribution results primarily from prompt gamma-ray emission due to the decay itself, which is simulated with \textsc{Pythia} 8.219~\cite{Sjostrand:2006za,Sjostrand:2007gs,Sjostrand:2014zea} including electroweak showering~\cite{Christiansen:2014kba} (see~\emph{e.g.}~\cite{Kachelriess:2007aj,Regis:2008ij,Mack:2008wu,Bell:2008ey,Dent:2008qy,Borriello:2008gy,Bertone:2008xr,Bell:2008vx,Cirelli:2009vg,Kachelriess:2009zy,Ciafaloni:2010ti}).  

These effects can be the only source of photons for channels such as $\chi \rightarrow \nu \bar{\nu}$.

  In addition, the electrons and positrons from these decays IC scatter off of cosmic background radiation (CBR), producing gamma-rays (see \emph{e.g.}~\cite{Murase:2015gea,Esmaili:2015xpa}).  The prompt contribution follows the spatial morphology obtained from the line-of-sight (LOS) integral of the DM density, which we model with a Navarro-Frenk-White~(NFW) profile~\cite{Navarro:1995iw,Navarro:1996gj}, setting the local DM density \mbox{$\rho = 0.3~\text{GeV}/\text{cm}^3$}, and the scale radius \mbox{$r_s = 20$ kpc} (variations to the profile lead to similar results, see the Supplementary Material).  We only consider IC scattering off of the cosmic microwave background~(CMB), as scattering from integrated stellar radiation and the infrared background is expected to be sub-dominant, see the Supplementary Material.
 For scattering off of the CMB, the resulting gamma-ray morphology also follows the LOS integral of the DM density.  Importantly, as scattering off of the other radiation fields only increases the gamma-ray flux, neglecting these effects is conservative.  In the same spirit, we conservatively assume that the electrons and positrons lose energy due to synchrotron emission in a rather strong, uniform $B = 2.0$ $\mu$G magnetic field (see \emph{e.g.}~\cite{Mao:2012hx,Haverkorn:2014jka,Beck:2014pma}) and show variations in the Supplementary Material.

In addition to the Galactic fluxes, there is an essentially isotropic extragalactic contribution, arising from DM decays throughout the broader Universe~\cite{Kribs:1996ac}. The extragalactic flux receives three important contributions: (1)~attenuated prompt emission; (2)~attenuated emission from IC of primary electrons and positrons; and~(3) emission from gamma-ray cascades. The cascade emission arises when an electron-positron pair is created by high-energy gamma rays scattering off of the CBR, inducing IC emission along with adiabatic energy loss.  We account for these effects following \cite{Murase:2012xs,Murase:2015gea}.

\vspace{-8pt}
\section{Data Analysis}
\label{sec:DataAnalysis}
\vspace{-10pt}
We assess how well predicted Galactic~(NFW-correlated) and extragalactic~(isotropic) fluxes describe the data using the profile-likelihood method (see {\it e.g.}~\cite{Rolke:2004mj}), described in more detail in the Supplementary Material.
To this end, we perform a template fitting analysis (using \texttt{NPTFit}~\cite{Mishra-Sharma:2016gis}) with 413 weeks of {\it Fermi} Pass 8 data collected from August 4, 2008 to July 7, 2016.  We restrict to the UltracleanVeto event class; furthermore, we only use the top quartile of events as ranked by point-spread function (PSF).  The UltracleanVeto event class is used to minimize contamination from cosmic rays, while the PSF cut is imposed to mitigate effects from mis-modeling bright regions.   We bin the data in 40 logarithmically-spaced energy bins between $200$ MeV and $2$ TeV, and we apply the recommended quality cuts \texttt{DATA\_QUAL>0 \&\& LAT\_CONFIG==1} and zenith angle less than $90^\circ$~\cite{fermi}.  The data is binned spatially using a HEALPix~\cite{Gorski:2004by} pixelation with \texttt{nside}=128.

We constrain this data to a region of interest~(ROI) defined by Galactic latitude $|b| \geq 20^\circ$ within $45^\circ$ of the Galactic Center~(GC). The Galactic plane is masked in order to avoid issues related to mismodeling of diffuse emission in that region.  Similarly, we do not extend our region out further from the GC to avoid over-subtraction issues that may arise when fitting diffuse templates over large regions of the sky (see \emph{e.g.}~\cite{Daylan:2014rsa,Linden:2016rcf,Narayanan:2016nzy}). Finally we mask all point sources~(PSs) in the 3FGL PS catalog~\cite{Acero:2015hja} at their 95\% containment radius.

Using this restricted dataset, we then 
independently fit templates in each energy bin in order to construct a likelihood profile as a function of the extragalactic and Galactic flux. 
We separate our model parameters into those of interest  ${\bf \psi}$ and the nuisance parameters ${\bf \lambda}$.  The ${\bf \psi}$ include parameters for an isotropic template to account for the extragalactic emission, along with a template following a LOS-integrated NFW profile to model the Galactic emission.  Note that both the prompt and IC contribute to the same template, see the Supplementary Material for justification. The ${\bf \lambda}$ include parameters for the flux from diffuse emission within the Milky Way, flux from the {\it Fermi} bubbles, flux from isotropic emission that does not arise from DM decay (\emph{e.g.} emission from blazars and other extragalactic sources, along with misidentified cosmic rays), and flux from PSs, both Galactic and extragalactic, in the 3FGL PS catalog. Importantly, each spatial template is given a separate, uncorrelated degree of freedom in the northern and southern hemispheres, further alleviating over-subtraction. 

In our main analysis, we use the Pass 7 diffuse model \texttt{gal\_2yearp7v6\_v0} (\texttt{p7v6}) to account for diffuse emission in the Milky Way, coming from gas-correlated emission (mostly pion decay and bremsstrahlung from high-energy electrons), IC emission, and emission from large-scale structures such as the {\it Fermi} bubbles~\cite{Su:2010qj} and Loop~1~\cite{Casandjian:2009}.  Additionally, even though the {\it Fermi} bubbles are included to some extent in the \texttt{p7v6} model, we add an additional degree of freedom for the bubbles, following the uniform spatial template given in~\cite{Su:2010qj}.
  We add a single template for all 3FGL PSs based on the spectra in~\cite{Acero:2015hja}, though we emphasize again that all PSs are masked at 95\% containment.  See the Supplementary Material for variations of these choices.

Given the templates described above, we are able to construct 2-d log-likelihood profiles $\log p_i(d_i | \{I^i_\text{iso}, I^i_\text{NFW}\})$ as functions of the isotropic and NFW-correlated DM-induced emission $I^i_\text{iso}$ and  $I^i_\text{NFW}$, respectively, in each of the energy bins $i$.  Here, $d_i$ is the data in that energy bin, which simply consists of the number of counts in each pixel.  The likelihood profiles are given by maximizing the Poisson likelihood functions over the $\lambda$ parameters.

Any decaying DM model may be constrained from the set of likelihood profiles in each energy bin, which are provided as Supplementary Data~\cite{supp-data}. Concretely, given a DM model ${\cal M}$, the total log-likelihood $\log p(d | {\cal M},  \{\tau, m_\chi\})$ is simply the sum of the $\log p_i$, where the intensities in each energy bin are functions of the DM mass and lifetime.  The test-statistics (TS) used to constrain the model is twice the difference between the log-likelihood at a given $\tau$ and the value at $\tau = \infty$, where the DM contributes no flux.
The 95\% limit is given by $\text{TS} = -2.71$.

In order to compare our gamma-ray results to potential signals from IceCube, we determine the region of parameter space where DM may contribute to the observed high-energy neutrino flux.  We use the recent high-energy astrophysical neutrino spectrum measurement by the IceCube collaboration~\cite{Aartsen:2015knd}.  In that work, neutrino flux measurements from a combination of muon-track and shower data are given in 9 logarithmically-spaced energy bins between 10 TeV and 10 PeV, under the assumption of equal flavor ratios and an isotropic flux.\footnote{Constraints at high masses may be improved by incorporating recent results from~\cite{Aartsen:2016ngq}, which focused on neutrino events with energies greater than 10 PeV.}  We assume that DM decays are the only source of high-energy neutrino flux.  In Fig.~\ref{Fig: bbResult} (assuming the DM decays exclusively to $b \bar b$) we show the region where the DM model provides at least a 3$\sigma$ improvement over the null hypothesis of no high-energy flux at all.  The best-fit point is marked with a star.   The blue region in Fig.~\ref{Fig: bbResult} is the best-fit region~\cite{Chianese:2016kpu} for explaining an apparent excess in the 2-year medium energy starting event (MESE) IceCube data, which extends down to energies $\sim$1 TeV~\cite{Aartsen:2014muf}.   

The dashed red curve, on other other hand, shows the 95\% limit that we obtain on this DM channel under the assumption that astrophysical sources also contribute to the high-energy flux.  We parameterize the astrophysical flux by a power-law with an exponential cut-off, and we marginalize over the slope of the power-law, the normalization, and the cut-off in order to obtain a likelihood profile for the DM model, as a function of $\tau$ and $m_\chi$.  We emphasize that we allow the spectral index to float, as opposed to the analysis of \cite{Chianese:2016kpu}, which fixes the index equal to two.

\vspace{-8pt}
\section{Interpretations}
\label{sec:Interp}
\vspace{-10pt}
In Fig.~\ref{Fig: bbResult}, we show our total constraint on the DM lifetime for a model where $\chi \to b\, \bar b$.  This result demonstrates tension in models where decaying DM explains or contributes to the astrophysical neutrino flux observed by IceCube.  PeV-scale decaying DM models have received attention recently (see \emph{e.g.}~\cite{Esmaili:2013gha,Feldstein:2013kka,Ema:2013nda,Zavala:2014dla,Bhattacharya:2014vwa,Higaki:2014dwa,Rott:2014kfa,Fong:2014bsa,Dudas:2014bca,Ema:2014ufa,Esmaili:2014rma,Murase:2015gea,Anchordoqui:2015lqa,Boucenna:2015tra,Ko:2015nma,Aisati:2015ova,Kistler:2015oae,Chianese:2016opp,Fiorentin:2016avj,Dev:2016qbd,DiBari:2016guw,Kalashev:2016cre,Chianese:2016smc}). 
In particular, while conventional astrophysical models such as those involving star-forming galaxies and galaxy clusters provide viable explanations for the neutrino data above 100\,TeV (see~\cite{Murase:2016gly} for a summary of recent ideas), the MESE data have been difficult to explain with conventional models~\cite{Murase:2015xka,Palladino:2016xsy}.
 Moreover, it is natural to expect heavy DM to slowly decay to the SM in a wide class of scenarios where, for example, the DM is stabilized through global symmetries in a hidden sector that are expected to be violated at the Planck scale or perhaps the scale of grand unification (the GUT scale).

From a purely data-driven point of view it is worthwhile to ask whether any set of SM final states may contribute significantly to or explain the IceCube data while being consistent with the gamma-ray constraints. 
In the Supplementary Material we provide limits on a variety of two-body SM final states.

 It is also important to interpret the bounds as constraints on the parameter space of UV models or gauge-invariant effective field theory~(EFT) realizations.  
If the decay is mediated by irrelevant operators, and given the long lifetimes we are probing, it is natural to assume very high cut-off scales $\Lambda$, such as the GUT scale $\sim$$10^{16} \text{\,GeV}$ or the Planck scale $m_\text{Pl} \simeq 2.4 \times 10^{18}$\,GeV.  We expect all gauge invariant operators connecting the dark sector to the SM to appear in the EFT suppressed by a scale $m_\text{Pl}$ or less (assuming no accidentally small coefficients and, perhaps, discrete global symmetries).  

It is also interesting to consider 
models that could yield signals relevant for this analysis.  Many cases are explored in the Supplementary Material, and here we highlight one simple option:  a hidden sector that consists of a confining gauge theory, at scale $\Lambda_{D}$~\cite{Faraggi:2000pv}, without additional light matter. Hidden gauge sectors that decouple from the SM at high scales appear to be generic in many string constructions (see~\cite{Halverson:2016nfq} for a recent discussion).  Denoting 
the hidden-sector field strength as $G_{D \mu \nu}$, then the lowest dimensional operator connecting the hidden sector to the SM appears at dimension-6:  $\mathcal{L} \supset \lambda_D\, {G_{D \mu \nu}\, G_D^{ \mu \nu}\, |H|^2 / \Lambda^2 }$, where $\lambda_D$ is a dimensionless coupling constant, $\Lambda$ is the scale where this operator is generated, and $H$ the SM Higgs doublet.  The lightest $0^{++}$ glueball state in the hidden gauge theory is a simple DM candidate $\chi$, with $m_\chi \sim \Lambda_{D}$, though heavier, long-lived states may also play important roles (see \emph{e.g.}~\cite{Forestell:2016qhc}).  The lowest dimension EFT operator connecting $\chi$ to the SM is then $\sim \chi\, |H|^2 \,\Lambda_{D}^3 / \Lambda^2$.  Furthermore, $\Lambda_{D}\gtrsim100\,$MeV in order to avoid constraints on DM self-interactions~\cite{Boddy:2014yra}.

\begin{figure}[t]
	\leavevmode
	\vspace{-.30cm}
	\begin{center}
        \includegraphics[width = 0.95 \columnwidth]{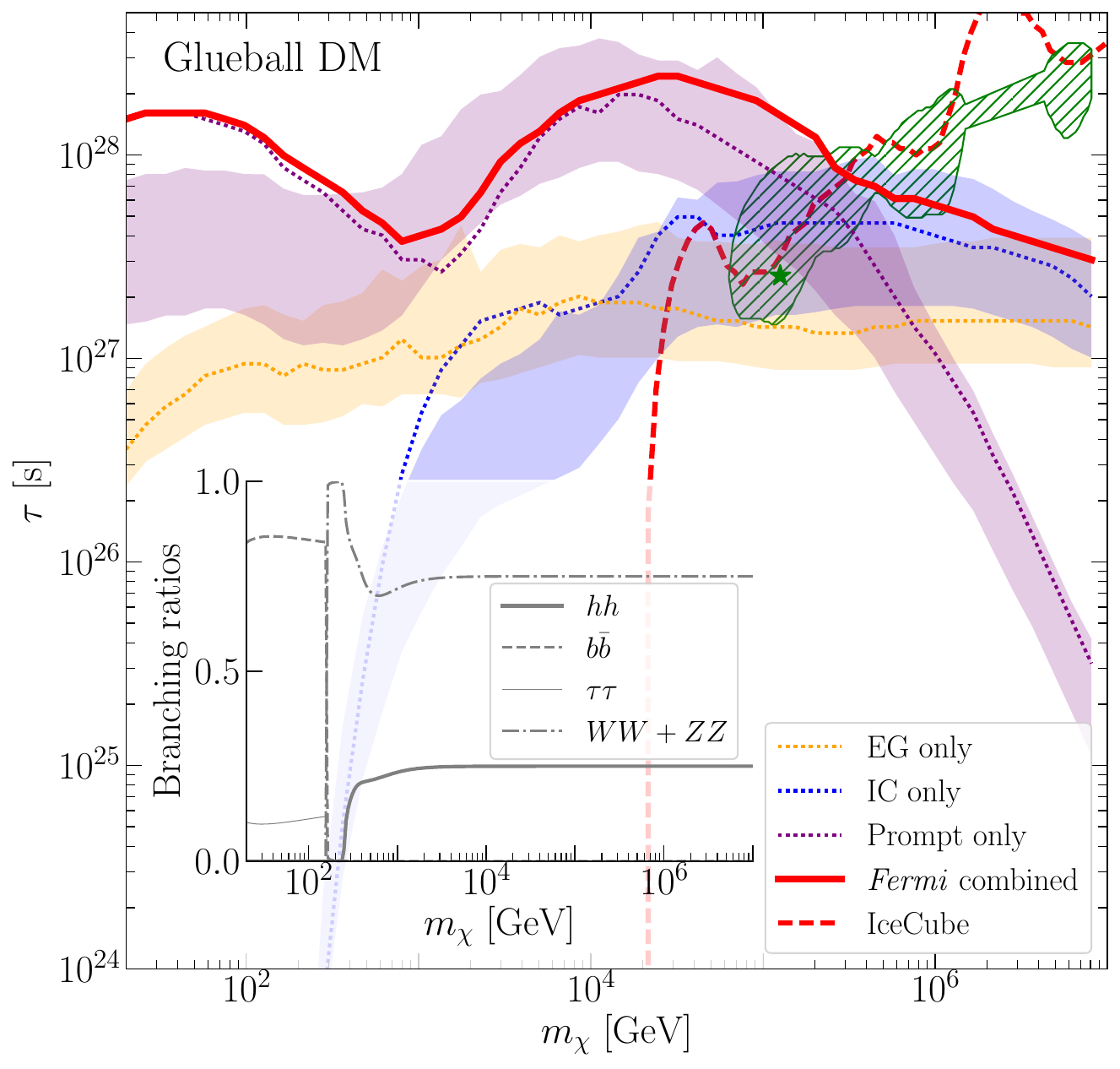}
	\end{center}
	\vspace{-.50cm}
	\caption{Limits  
	on decaying glueball DM (see text for detals).
	We show limits obtained from prompt, IC, and EG emission only, along with the 95\% confidence window for the expectation of each limit from MC simulations. 
	 Furthermore, the parameter space where the IceCube data may be interpreted as a $\sim$$3\sigma$ hint for DM is shown in shaded green, with the best fit point represented by the star. 
	 (inset) The dominant glueball DM branching ratios.
	}
	\vspace{-0.15in}
	\label{Fig: glue}
\end{figure}

At masses comparable to and lower than the electroweak scale, the glueball decays primary to $b$ quarks through mixing with the SM Higgs, while at high masses the glueball decays predominantly to $W^\pm$, $Z^0$, and Higgs boson pairs (see the inset of Fig.~\ref{Fig: glue} for the dominant branching ratios).  In the high-mass limit, the lifetime is approximately 
 \es{tau_glue}{
\!\!\tau &\simeq 5 \cdot 10^{27}  \, \text{s} \left( { 3 \over N_D} {1 \over 4\, \pi  \lambda_D} \right)^2 \left( {\Lambda \over m_\text{Pl} } \right)^4 \left( {0.1  \, \text{PeV}  \over \Lambda_{D} } \right)^5 ,
 }
with $N_D$ the number of colors.  This is roughly the right lifetime to be relevant for the IceCube neutrino flux.    
 
   In Fig.~\ref{Fig: glue}, we show our constraint on this glueball model.  Using Eq.~\eqref{tau_glue}, these results suggest that models with $\Lambda_{D} \gtrsim 0.1 \, \, \text{PeV}$, $\lambda_D \gtrsim 1/(4 \pi)$, and $\Lambda = m_\text{Pl}$ are excluded.   As in Fig.~\ref{Fig: bbResult}, the shaded green is the region of parameter space where the model may contribute significantly to IceCube, and the dashed red line provides the limit we obtain from IceCube allowing for an astrophysical contribution to the flux.  As in the case of the $b\, \bar b$ final state, the gamma-ray limits derived in this work are in tension with the decaying-DM origin of the signal. 

Figure~\ref{Fig: glue} also illustrates the relative contribution of prompt, IC and extragalactic emissions to the total limit. The 95\% confidence interval is shown for each source, assuming background templates only, where the normalizations are fit to the data.
Across almost all of the mass range, and particularly at the highest masses, the limits obtained on the real data align with the expectations from MC.  In the statistics-dominated regime, we would expect the real-data limits to be consistent with those from MC, while in the systematics dominated regime the limits on real data may differ from those obtained from MC.  This is because the real data can have residuals coming from mis-modeling the background templates, and the overall goodness of fit may increase with flux from the NFW-correlated template, for example, even in the absence of DM.  Alternatively, the background templates may overpredict the flux at certain regions of the sky, leading to over-subtraction issues that could make the limits artificially strong.

\vspace{-8pt}
\section{Discussion}
\label{sec:Discussion}
\vspace{-10pt}
In this work, we presented some of the strongest limits to date on decaying DM from a dedicated analysis of {\it Fermi} gamma-ray data incorporating spectral and spatial information, along with up-to-date modeling of diffuse emission in the Milky Way.  Our results disfavor a decaying DM explanation of the IceCube high-energy neutrino data.

There are several ways that our analysis could be expanded upon.
We have not attempted to characterize the spectral composition of the astrophysical contributions to the isotropic emission, which may strengthen our limits.  On the other hand, ideally, for a given, fixed decaying DM flux in the profile likelihood, we should marginalize not just over the normalization of the diffuse template but also over all of the individual components that go into making this template, such as IC emission and bremsstrahlung.

A variety of strategies beyond those described here have been used to constrain DM lifetimes (see \emph{e.g.}~\cite{Ibarra:2013cra} for a review).  These include gamma-ray line searches, such as those performed in~\cite{Abdo:2010nc,Vertongen:2011mu,Ackermann:2012qk,Ackermann:2013uma}, which are complementary to the constraints on broader energy emission given in this Letter. Limits from direct decay into neutrinos have also been considered \cite{Esmaili:2012us}. Less competitive limits  
have been set on DM decays resulting in broad energy deposition and nearby galaxies and galaxy clusters~\cite{Dugger:2010ys,Huang:2011xr}, large scale Galactic and extragalactic emission~\cite{Cirelli:2009dv,Zhang:2009ut,Zaharijas:2010ca,Ackermann:2012rg,Zaharijas:2012dr}, Milky Way Dwarfs~\cite{Aliu:2012ga,Baring:2015sza}, and the CMB~\cite{Slatyer:2016qyl}.  The upcoming Cherenkov Telescope Array~(CTA) experiment~\cite{Consortium:2010bc}
may have similar sensitivity as our results to DM masses $\sim$10 TeV~\cite{Pierre:2014tra}.  However, more work needs to be done in order to assess the potential for CTA to constrain or detect heavier, $\sim$PeV decaying DM.  
On the other hand, the High-Altitude Walter Cherenkov Observatory~(HAWC)~\cite{Abeysekara:2013tza} and air-shower experiments such as Tibet AS+MD~\cite{Sako:2009xa} will provide meaningful constraints on the Galactic diffuse gamma-ray emission. The constraints on DM lifetimes might be as stringent as ${10}^{27}-{10}^{28}$~s for PeV masses and hadronic channels, assuming no astrophysical emission is seen~\cite{Ahlers:2013xia,Murase:2015gea,Esmaili:2015xpa}.

Finally, we mention that our results also have implications for possible decaying DM interpretations (see {\it e.g.}~\cite{Cheng:2016slx}) of the positron~\cite{Aguilar:2013qda,Ting:2238506} and antiproton fluxes~\cite{ams02pos} measured by AMS-02. 
Recent measurements of the positron flux appear to exhibit a break at high masses that could indicate evidence for decaying DM to, for example, $e^+ \,e^-$ with $m_\chi \sim$ 1 TeV and $\tau \sim$ $10^{27} \text{ s}$.  However, our results appear to rule out the decaying DM interpretation of the positron flux for this and other final states.  For example, in the $e^+\, e^-$ case our limit for $m_\chi \sim$ 1 TeV DM is $\tau \gtrsim 5 \times 10^{28} \text{ s}$.

\vspace{-8pt}
\section*{Acknowledgements}
\vspace{-10pt}
We thank John Beacom, Keith Bechtol, Kfir Blum, Jim Cline, Jonathan Cornell, Arman Esmaili, Andrew Fowlie, Benjamin Fuks, Philip Ilten, Koji Ishiwata, Joachim Kopp, Hongwan Liu, Ian Low, Naoko Kurahashi-Neilson, Farinaldo Queiroz, Tracy Slatyer, Yue-Lin Sming Tsai, and Christoph Weniger for useful conversations. We also thank Lars Mohrmann for providing us with the IceCube data from the maximum likelihood analysis.
TC is supported by an LHC Theory Initiative Postdoctoral Fellowship, under the National Science Foundation grant PHY-0969510. The work of KM is supported by NSF Grant No. PHY-1620777. NLR is supported in part by the American Australian Association's ConocoPhillips Fellowship. BRS is supported by a Pappalardo Fellowship in Physics at MIT.  The work of BRS was performed in part at the Aspen Center for Physics, which is supported by National Science Foundation grant PHY-1066293. This work is supported by the U.S. Department of Energy~(DOE) under cooperative research agreement DE-SC-0012567 and DE-SC-0013999.

\twocolumngrid
\vspace{-8pt}
\section*{Bibliography}
\vspace{-10pt}
\def\bibsection{}
\bibliographystyle{utphys}
\bibliography{DMDecay}

\clearpage
\newpage
\maketitle
\onecolumngrid
\begin{center}
\textbf{\large Gamma-ray Constraints on Decaying Dark Matter and Implications for IceCube  } \\ 
\vspace{0.05in}
{ \it \large Supplementary Material}\\ 
\vspace{0.05in}
{ Timothy Cohen, Kohta Murase, Nicholas L. Rodd, Benjamin R. Safdi, and Yotam Soreq}
\end{center}
\onecolumngrid
\setcounter{equation}{0}
\setcounter{figure}{0}
\setcounter{table}{0}
\setcounter{section}{0}
\setcounter{page}{1}
\makeatletter
\renewcommand{\theequation}{S\arabic{equation}}
\renewcommand{\thefigure}{S\arabic{figure}}
\renewcommand{\thetable}{S\arabic{table}}
\newcommand\ptwiddle[1]{\mathord{\mathop{#1}\limits^{\scriptscriptstyle(\sim)}}}

The supplementary material is organized as follows.  In Sec.~\ref{sec: methods}, we provide more detail regarding the methods used in the main body of this work.  In particular, we discuss the calculations of the gamma-ray spectra and the data analysis.  In Sec.~\ref{sec: detailed results}, we give extended results beyond those given in the main body.  Then, in Sec.~\ref{sec: systematics}, we characterize and test sources of systematic uncertainty that could affect our results.  Lastly, in Sec.~\ref{sec: models}, we overview EFTs for decaying DM and give constraints on explicit models beyond those discussed in the main text.

\section{Methods}
\label{sec: methods}

We begin this section by detailing the calculations of the prompt and secondary spectra from DM decay.  Then, we discuss in detail the likelihood profile technique used in this paper.  

\subsection{Spectra}

This section provides a more detailed description of the gamma-ray spectra that result from heavy DM decay.  There is a natural decomposition into three components: (1)~prompt Galactic gamma-ray emission, (2)~Galactic inverse Compton~(IC) emission from high-energy electrons and positrons up-scattering background photons, and (3)~extragalactic flux from DM decay outside of our Galaxy.  As mentioned in the main text, when calculating the prompt spectrum (and the primary electron and positron flux) it is crucial, for certain final states, to included electroweak radiative processes, as these may be the only source of gamma-ray emission.  To illustrate this point, in Tab.~\ref{table:Particles} we show the average number of primary gamma-rays, neutrinos, and electrons and their energy fraction coming from DM decay to $b \,\bar b$ and $\nu\, \bar \nu$ for various DM masses. 
We note that for $m_\chi=100\,$GeV there are in average 3\,(0) hadrons in the final state, while for $m_\chi=1\,$PeV there are  77\,(1) hadrons for the $b\bar{b}\,(\nu_e\bar{\nu}_e)$ decay mode. The energy fraction of these hadrons is 13\,(0)\,\% and 16\,(0.5)\,\% for $b\bar{b}\,(\nu_e\bar{\nu}_e)$ modes with a DM mass of 100\,GeV and 1\,PeV, respectively.  In addition, the energy fractions of photons, neutrinos and electrons are almost independent of the DM mass for the $b\bar{b}$ decay mode. This can be understood as the majority of these final states originate from pion decays.
{
\begin{table}[h]
\renewcommand{\arraystretch}{2}
\setlength{\arrayrulewidth}{.3mm}
\centering
\setlength{\tabcolsep}{0.8em}
\begin{tabular}{ c  c | c | c | c | c | c || c | c | c | c | c |}
\cline{3-12}
& & \multicolumn{5}{ c|| }{$\chi \to b \,\bar{b}$} & \multicolumn{5}{ c| }{$\chi \to \nu_e\, \bar{\nu}_e$} \\ 
\cline{1-1} \cline{3-12}
\multicolumn{1}{|c|}{$m_{\chi}$} & & $\gamma$ & $\nu$ & $e^-/e^+$ & $W^\pm/Z^0$\,\footnote{Unlike the other particles in this table, the $W^\pm$ and $Z^0$ are unstable and so we only count them at the intermediate stage of the cascade.}  & All & $\gamma$ & $\nu$ & $e^-/e^+$ & $W^\pm/Z^{0\,a}$ & All \\ \hline \hline
\multicolumn{1}{|c|}{\multirow{2}{*}{\rotatebox[origin=c]{90}{100 GeV}} }& \# of particles & 26 & 66 & 23 & 0 & 120 & 0 & 2 & 0 & 0 & 2 \\
\multicolumn{1}{|c|}{} & energy fraction & 0.27 & 0.44 & 0.17 & 0 & - & 0 & 1 & 0 & 0  & -  \\ \hline
\multicolumn{1}{|c|}{\multirow{2}{*}{\rotatebox[origin=c]{90}{1 TeV}}}& \# of particles & 58& 150 & 51 & 0.006 & 270 & 0.37 & 3 & 0.36 & 0.026 & 3.8 \\
\multicolumn{1}{|c|}{} & energy fraction & 0.28 & 0.44 & 0.17 & 0.002 & - & 0.001 & 0.99 & 0.007 & 0.006 & - \\ \hline
\multicolumn{1}{|c|}{\multirow{2}{*}{\rotatebox[origin=c]{90}{10 TeV}}}& \# of particles & 120 & 320 & 110 & 0.039 & 570 & 2.0 & 7.4 & 1.9 & 0.14 & 12 \\
\multicolumn{1}{|c|}{}& energy fraction & 0.28 & 0.44 & 0.17 & 0.006 & - & 0.004 & 0.96 & 0.034 & 0.020 & - \\ \hline
\multicolumn{1}{|c|}{\multirow{2}{*}{\rotatebox[origin=c]{90}{100 TeV}}}& \# of particles & 250 & 660 & 230 & 0.098 & 1200 & 5.1 & 15 & 4.8 & 0.35 & 26 \\
\multicolumn{1}{|c|}{} & energy fraction & 0.29 & 0.44 & 0.17 & 0.009 & -  & 0.007 & 0.91 & 0.078  & 0.033  & -\\ \hline
\multicolumn{1}{|c|}{\multirow{2}{*}{\rotatebox[origin=c]{90}{1 PeV}}}& \# of particles & 490 & 1300 & 440 & 0.18 & 2300 & 9.2 & 27 & 8.7 & 0.64 & 46 \\
\multicolumn{1}{|c|}{} & energy fraction & 0.29 & 0.44 & 0.18 & 0.013 & - & 0.009 & 0.86 & 0.13 & 0.045 & - \\ \hline
\end{tabular}
\caption{Average number of final state particles (upper line) and their average energy fraction (lower line) for DM decay to bottom quarks or electron neutrinos. For the neutrino case, the presence of electroweak corrections has a large impact on the resulting spectrum for higher masses, whereas for the hadronic final state the effect is less important.
}
\label{table:Particles}
\end{table}
}

\begin{figure}[t]
	\leavevmode
	\begin{center}$
	\begin{array}{cc}
	\scalebox{0.5}{\includegraphics{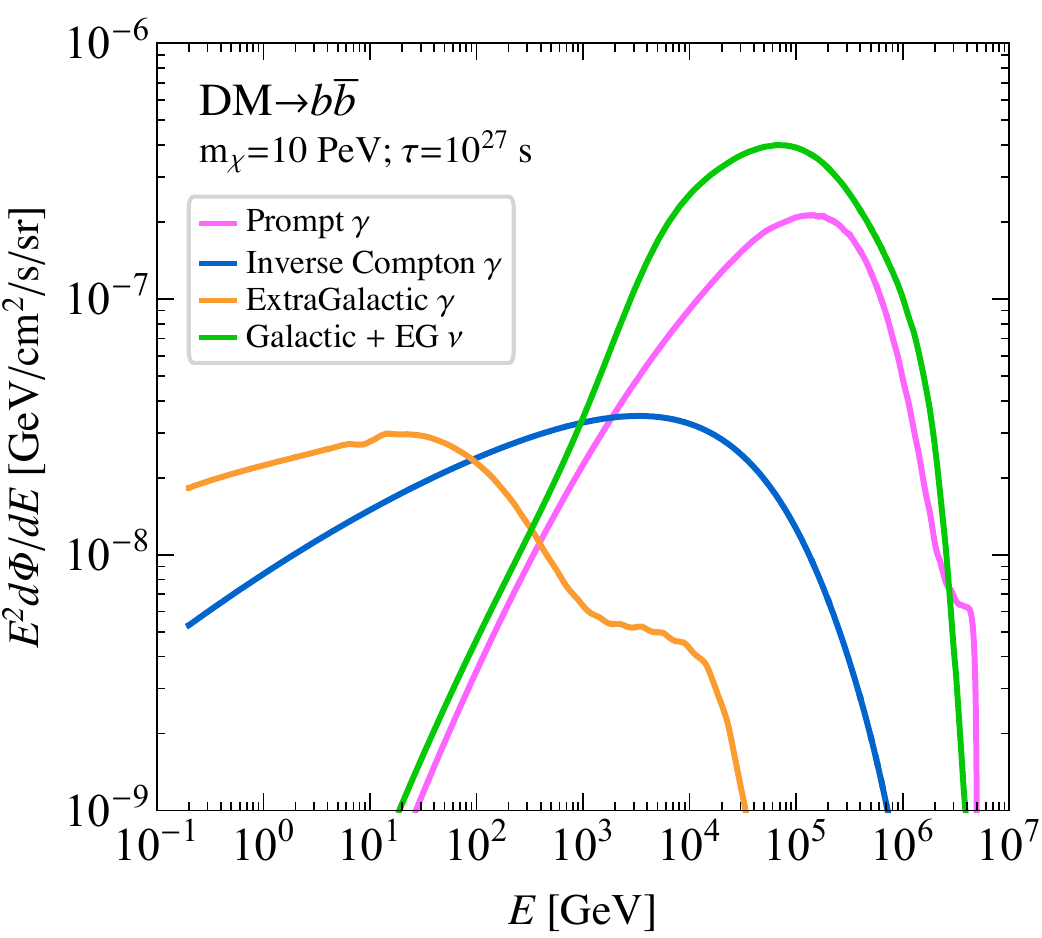}} &
	\scalebox{0.5}{\includegraphics{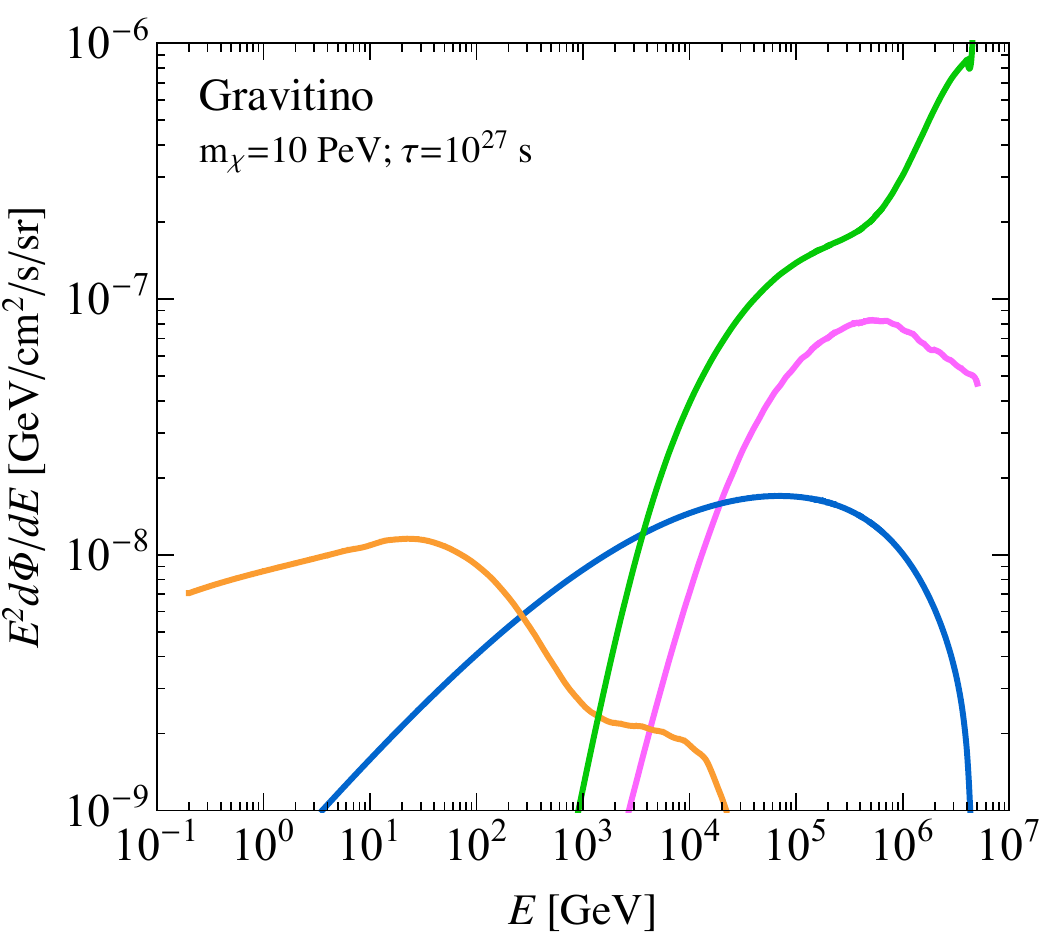}}
	\end{array}$
	\end{center}
	\vspace{-.70cm}
	\caption{Gamma-ray and neutrino spectra for DM decaying to $b\,\bar{b}$ (left) and a model of gravitino DM (right) as detailed in Sec.~\ref{sec: models} below, with $m_\chi = 10$ PeV and $\tau = 10^{27}$ s.  All fluxes are normalized within the ROI used in our main analysis. {\it Fermi} can detect photons in the range $\sim 0.2 - 2000$\,GeV. For heavy DM decays, the flux in the {\it Fermi} energy range is dominated by the IC and extragalactic contributions, rather than the prompt Galactic emission.
	}
	\vspace{-0.15in}
	\label{Fig: BasicSpec}
\end{figure}

\begin{figure}[b]
	\leavevmode
	\begin{center}$
	\begin{array}{cc}
	\scalebox{0.5}{\includegraphics{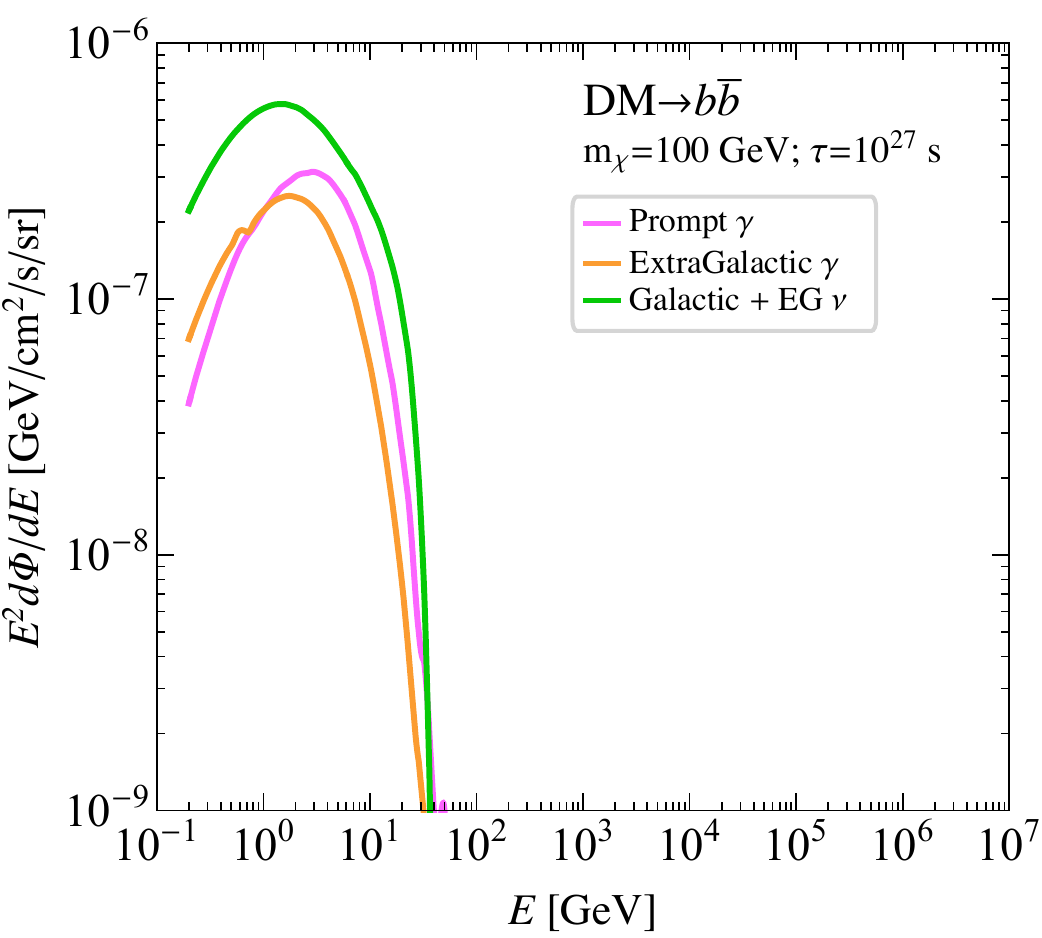}}  &
	\scalebox{0.5}{\includegraphics{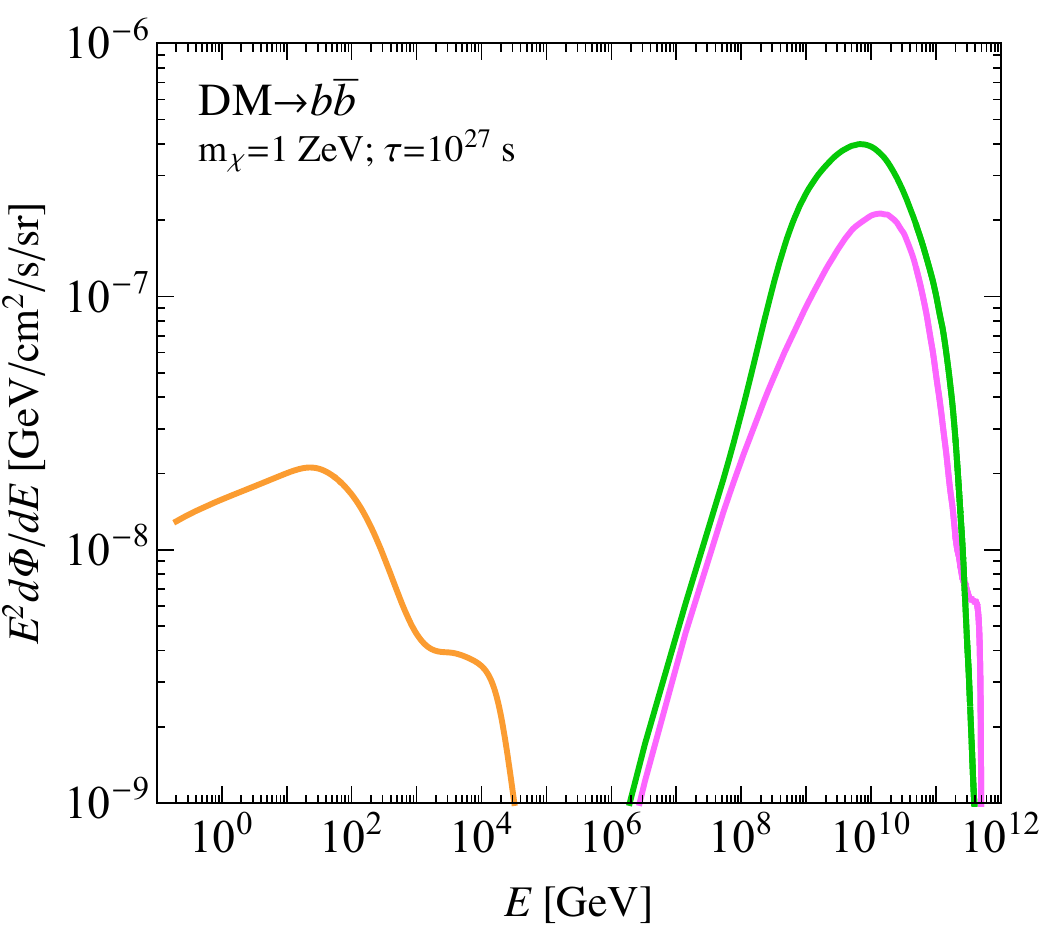}}
	\end{array}$
	\end{center}
	\vspace{-.70cm}
	\caption{Gamma-ray and neutrino spectra for DM decaying to $b\,\bar{b}$ for two different DM masses: $100$ GeV (left) and $10^{12}$ GeV (right). These should be compared to the {\it Fermi} energy range of $\sim 0.2 - 2000$ GeV. For the lighter DM case, prompt emission dominates, whilst at higher masses the dominant contribution is from the extragalactic flux. In neither of these cases is IC emission relevant, this only contributes meaningfully for intermediate $\mathcal{O}({\rm PeV-EeV})$ masses, as seen on the left of Fig.~\ref{Fig: BasicSpec}.
	}
	\vspace{-0.15in}
	\label{Fig: highlowbSpec}
\end{figure}

Additionally, we show the typical number of radiated $W$ and $Z$ bosons.  In the $b \bar b$ case, electroweak corrections are not significant even for 1\,PeV DM.  However, in the $\nu \bar \nu$ case the radiated $W$ and $Z$ bosons are responsible for the majority of the primary particles (and all of the gamma-rays and electrons) at masses above the electroweak scale. The importance of these electroweak corrections on dark matter annihilation/decay spectra have been previously noted (see~\emph{e.g.}~\cite{Kachelriess:2007aj,Regis:2008ij,Mack:2008wu,Bell:2008ey,Dent:2008qy,Borriello:2008gy,Bertone:2008xr,Bell:2008vx,Cirelli:2009vg,Kachelriess:2009zy,Ciafaloni:2010ti}). For DM masses above 10\,PeV, the large number of final states implies that generation of the spectra through showering in \textsc{Pythia} is no longer practical. We discuss in Appendix~\ref{sec:ExtendingToGUTScale} how we extend our spectra beyond these masses.\footnote{Publicly available DM spectra, such as those in~\cite{Cirelli:2010xx,Elor:2015tva,Elor:2015bho}, do not extend up to these high masses, which is why we have recalculated them. While there are certainly modeling errors associated with running \textsc{Pythia} at these very high energies, they are expected to be subdominant to the astrophysical uncertainties inherent in this analysis. We extend the spectra above 10 PeV by rescaling the appropriately normalized spectrum, as described and validated in the Supplementary Material.}

As was shown in Fig.~\ref{Fig: glue} in the main text, the prompt flux tends to be most important for lower DM masses near the {\it Fermi} energy range, while the IC emission may play a leading role for DM masses near the PeV scale.  The extragalactic flux is important over the whole mass range, but at very high masses -- well above the PeV scale -- the extragalactic flux may be the only source of gamma-rays in the {\it Fermi} energy range.  To illustrate these points, Fig.~\ref{Fig: BasicSpec} shows the gamma-ray and neutrino spectra at Earth, normalized to within the ROI used in our main analysis, for 10\,PeV DM decay with $\tau = 10^{27}$ s.  We consider two final states, $b \,\bar b$ (left) and the gravitino model (right), which is described in more detail later in this Supplementary Material.   

\begin{figure}[t]
	\leavevmode
	\begin{center}$
	\begin{array}{cc}
	\scalebox{0.5}{\includegraphics{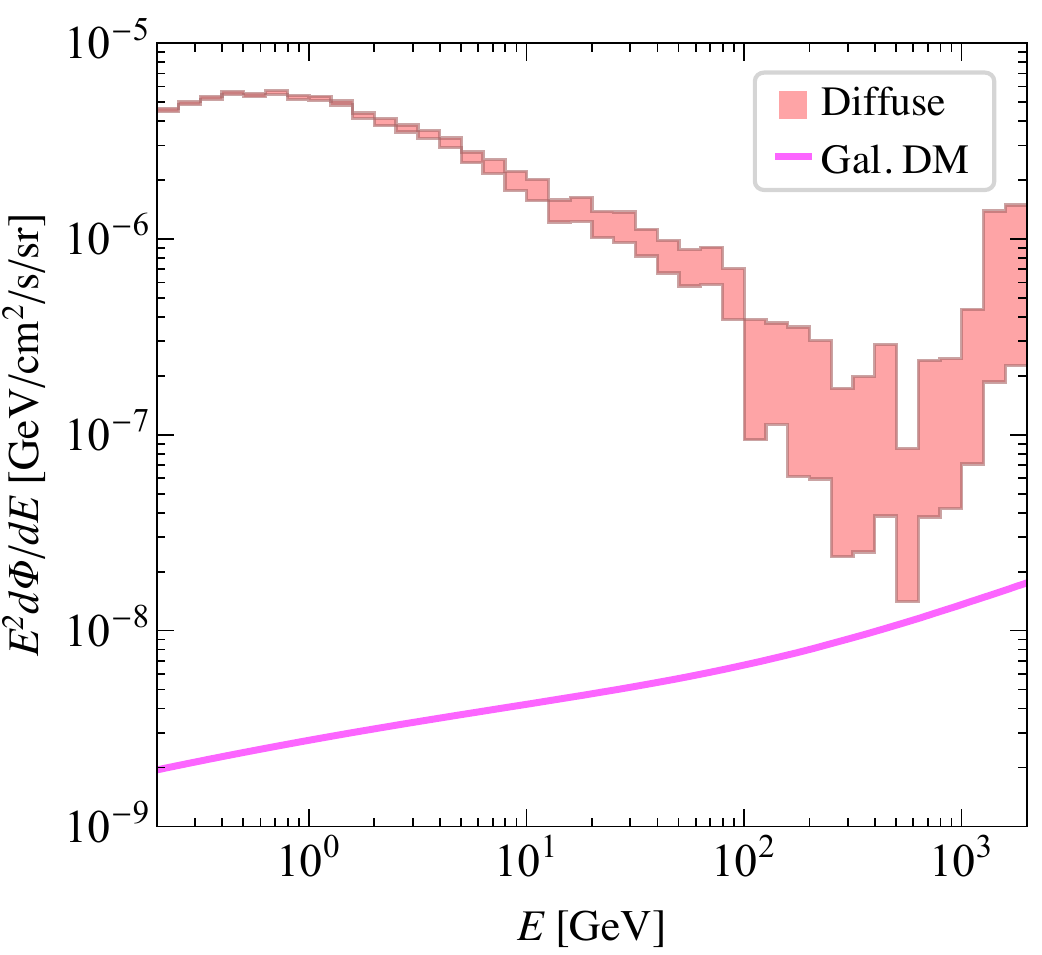}}  &
	\scalebox{0.5}{\includegraphics{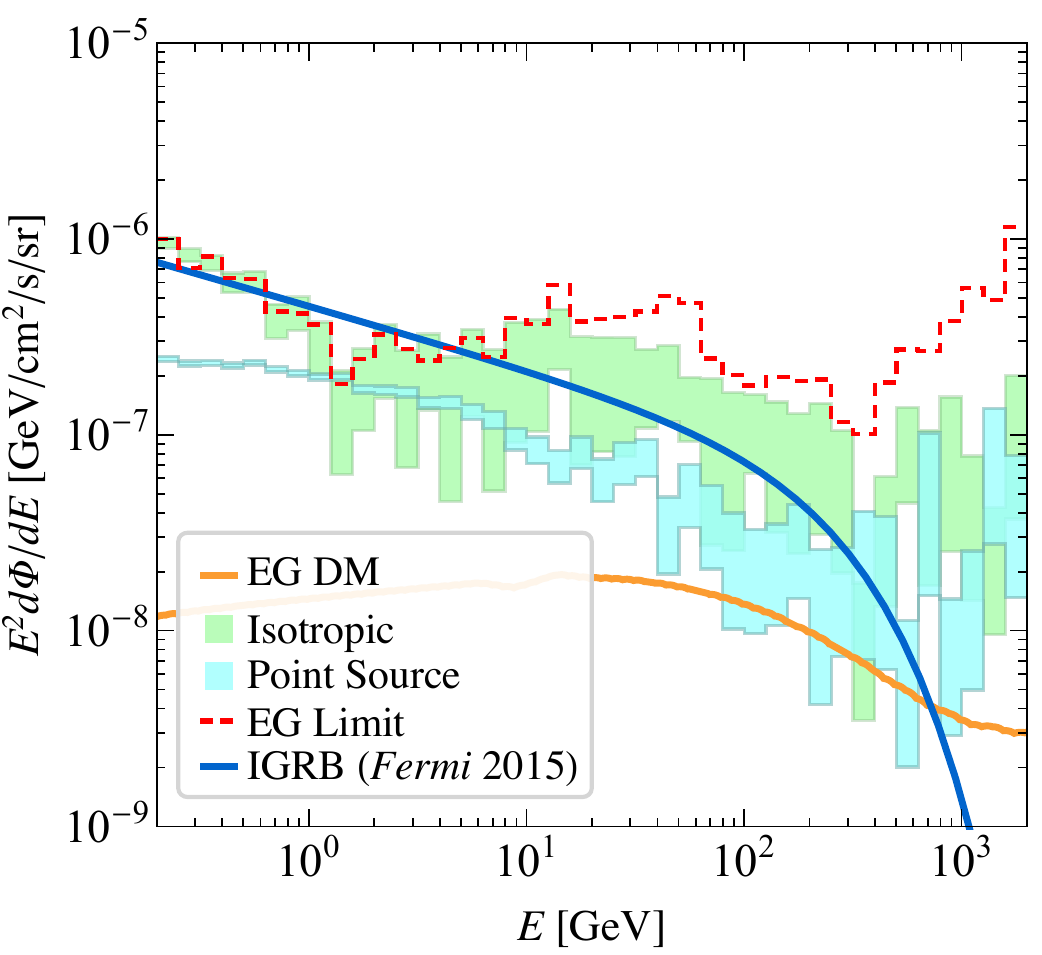}}
	\end{array}$
	\end{center}
	\vspace{-.70cm}
	\caption{A comparison between the $1$ PeV $b\, \bar{b}$ DM spectrum and that from our background models, for the largest lifetime we can constrain using only either Galactic ($\tau = 8.4 \times 10^{27}$ s) or extragalactic ($\tau = 1.7 \times 10^{27}$ s) DM flux. Spectra are averaged over the ROI used in our analysis. Left: Here we show the diffuse Galactic spectrum, compared to the smallest Galactic (prompt and IC) flux we can constrain. For the diffuse model we show the 68\% confidence interval determined from the posterior of our fit in each energy bin. Diffuse emission is responsible for the vast majority of the photons seen in our analysis, and it sits several orders of magnitude above the DM flux we can constrain in most energy bins. Right: The 68\% confidence intervals on the spectrum of our isotropic and point source models, compared to the weakest extragalactic DM flux we can constrain.  We also show in this plot the bin-by-bin 95\% limit we set on extragalactic flux, homogenious across the northern and southern sub-regions. Further, we illustrate the IGRB as measured by {\it Fermi} \cite{Ackermann:2014usa}, which is in good agreement with our isotropic spectrum across most of the energy range. See text for details. }
	\vspace{-0.15in}
	\label{Fig: DMvsBkgSpec}
\end{figure}

Importantly, for DM masses  $\gtrsim$1\,TeV, the gravitino decays roughly 50\% of the time into $W^{\pm}\, \ell^{\mp}$, where $\ell^{\mp}$ are SM leptons, and 50\% of the time into $Z^0\, \nu$ and $h \, \nu$.  These latter two final states are responsible for the sharp rise in the Galactic and extragalactic neutrino spectrum in the gravitino model at energies approaching the DM mass (10\,PeV in this case).  In both cases, however, the prompt gamma-ray spectra are seen to be sub-dominant within the {\it Fermi} energy range, which extends up to $\sim$2\,TeV.  At the upper end of the {\it Fermi} energy range, the IC emission is the dominant source of flux, while the extragalactic emission extends to much lower energies. 

To illustrate this point further, we show in Fig.~\ref{Fig: highlowbSpec} the $b\, \bar b$ final-state spectra for $m_\chi=100$\,GeV and $1$\,ZeV $\big(=10^{12} \text{ GeV}\big)$.  In the low-mass case, the IC emission is produced in the Thomson regime and peaks well below the {\it Fermi} energy range. Furthermore, in this case the extragalactic spectrum is generally sub-dominant to the prompt Galactic emission. In the high-mass case, the extragalactic flux is the only source of emission within the {\it Fermi} energy range. Indeed, it is well known that the extragalactic spectrum approaches a universal form, regardless of the primary spectra (\emph{e.g.} see~\cite{Murase:2012xs}; also as plotted in Fig.~\ref{Fig:UniversalSpec}). This can be seen by comparing the extragalactic spectrum on the right of Fig.~\ref{Fig: highlowbSpec} to that on the left on Fig.~\ref{Fig: BasicSpec}, and this is explored in more detail in Sec.~\ref{sec:ExtendingToGUTScale}. Finally for the ZeV DM decays, the IC emission is still largely peaked in the {\it Fermi} energy range, but has now transitioned completely to the Klein-Nishina regime, where the cross section is greatly reduced. As such its contribution is several orders of magnitude sub-dominat to the extragalactic flux. Note that in Fig.~\ref{Fig: highlowbSpec}, and in subsequent spectral plots, we have used a galactic $J$-factor that is averaged over our ROI. In detail, if we define $\rho(s,l,b)$ to be the DM density as a function of distance from Earth $s$, as well as galactic longitude $l$ and latitude $b$, then we used:
\begin{equation}
J = \int_{\rm ROI} \text{d} \Omega \int \text{d}s\, \rho(s,l,b) / \int_{\rm ROI} \text{d} \Omega \simeq 4.108 \times 10^{22}~{\rm GeV}~{\rm cm}^{-2}\,.
\end{equation}
This is larger than the all-sky averaged value by a factor of 2.6.

In the main text, we assumed that for the energies relevant for {\it Fermi}, the IC morphology will be effectively identical to that of the prompt DM decay flux. This justified the combination of the prompt and IC flux into a single spatial template which followed the above $J$-factor. 
In principle there are at least three places additional spatial dependence could enter, beyond the prompt $e^{\pm}$ spatial distribution injected by DM decays: 1. the distribution of the seed photon fields; 2. the distribution of the magnetic fields under which the electrons cool; and 3. the diffusion of the $e^{\pm}$. 
Referring to the first of these, there are three fields available to up-scatter off: the CMB, the integrated stellar radiation, and the infrared background due to the irradiated stellar radiation. These last two are position dependent and tend to decrease rapidly off the plane. So as long as we look off the plane, as we do, the CMB dominates and is position independent. Importantly, neglecting the other contributions is conservative, as they would only contribute additional flux. 
Regarding the second point, the regular and halo magnetic fields play an important role in the $e^{\pm}$ cooling. The former component highly depends on the Galactic latitude and decays off the plane; it is subdominant to the halo magnetic field in our ROI, so we ignore it.  
Finally, for the energies of interest, the diffusion of the $e^{\pm}$ can be neglected to a good approximation on the scales of interest, as discussed in \cite{Esmaili:2015xpa}. The halo field is expected to be strong enough for electrons and positrons to lose their energy in the halo.

Finally, Fig.~\ref{Fig: DMvsBkgSpec} shows the spectrum of the weakest Galactic and extragalactic DM fluxes we can constrain for $1$\,PeV DM decaying to bottom quarks, directly compared to the background contributions. In these figures, the three background components from our fits -- diffuse, isotropic, and point source emission -- are shown via a band between the 16 and 84 percentiles on these parameters extracted from the posterior, where the values are given directly in each of our 40 energy bins. Between these figures we see that diffuse emission dominates over essentially the entire energy range. We also see that the value of the isotropic flux is not particularly well constrained within our small ROI, especially at higher energy.  It is is important to note that our isotropic spectrum is found by averaging the spectra in the north and south, which are fit independently.  As a comparison, we also show the 95\% limit on homogenous extragalactic emission, which is by definition the same in the northern and southern hemispheres.  Reassuringly, our limit on extragalactic emission tends to be weaker than the isotropic gamma-ray background (IGRB) as measured by {\it Fermi} \cite{Ackermann:2014usa}, which is also shown in that figure.  The IGRB was determined from a dedicated analysis at high-latitudes using a data set with very low cosmic-ray contamination.  Even though our ROI and data set are far from ideal for determining the IGRB, we see that our isotropic spectrum is generally in very good agreement with the {\it Fermi} IGRB up to energies of around a few hundred GeV; at higher energies, our isotropic spectrum appears higher than the IGRB, perhaps because of cosmic ray contamination.  However, this should only make our high-energy extragalactic results conservative.

\subsection{Data analysis}

\begin{figure}[b]
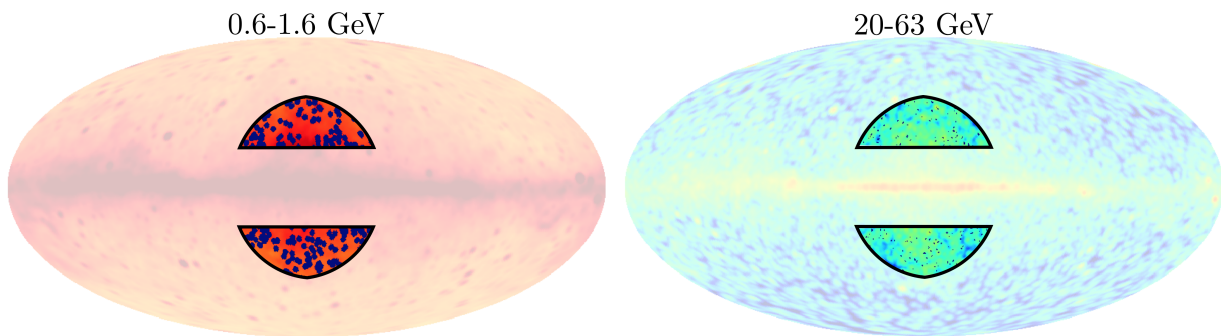

	\leavevmode
	\begin{center}$
	\begin{array}{cc}
	\scalebox{0.22}{\includegraphics{plots/ROI_LowE.pdf}} &
	\scalebox{0.22}{\includegraphics{plots/ROI_HighE.pdf}}
	\end{array}$
	\end{center}
	\vspace{-.40cm}
	\caption{The data within our Region of Interest (ROI), defined by $|b| > 20^{\circ}$ and $r < 45^{\circ}$, where $r$ is the angular distance from the GC. This ROI is shown in the context of the full data, shown with a lower opacity, for two different energy ranges: 0.6-1.6 GeV (left) and 20-63 GeV (right). In both cases the data has been smoothed to $2^{\circ}$, and all 3FGL point sources within our ROI have been masked at their 95\% containment radius. These are shown in blue, and are much larger on the left than the right as the {\it Fermi} PSF increases with decreasing energy. In our lowest energy bin (not shown), the point source mask covers most of our ROI.  In both figures, red shades indicate increased photon counts, while in the left (right) orange (blue) shades illustrate regions of low photon counts.}
	\vspace{-0.15in}
	\label{Fig:ROI}
\end{figure}
 
In this section, we expand upon the profile-likelihood analysis technique used in this work (see~\cite{Rolke:2004mj} for comments on this method). The starting point for this is the data itself, which we show in Fig.~\ref{Fig:ROI}. There we show our ROI in the context of the full dataset. Recall this ROI is defined by $|b| > 20^{\circ}$ and $r < 45^{\circ}$, with 3FGL PSs masked; this particular choice is discussed in detail in Sec.~\ref{sec: systematics}. The raw {\it Fermi} data is a list of photons with associated energies and positions on the sky. We bin these photons into 40 energy bins, indexed by $i$, that are equally log spaced from 200 MeV and 2 TeV. In each energy bin we then take the resulting data $d_i$, and spatially bin it using a HEALPix~\cite{Gorski:2004by} pixelation with \texttt{nside}=128. This divides our ROI into 12,474 pixels (before the application of a point source mask), which we index with $p$. The result of this energy and spatial binning reduces the raw data into a list of integers $n^p_i$ for the number of photons in pixel $p$ in the $i^\text{th}$ energy bin.

To determine the allowable DM decay contribution to this data, we need to describe it with a set of model parameters ${\bf \theta} = \{ {\bf \psi}, {\bf \lambda} \}$. As discussed in the main text, ${\bf \psi}$ are the parameters of interest which describe the DM flux, while ${\bf \lambda}$ are the set of nuisance parameters. In detail ${\bf \psi}$ accounts for the Galactic and separately extragalactic DM decay flux, and ${\bf \lambda}$ models the Galactic diffuse emission, {\it Fermi} bubbles, isotropic flux, and emission from PSs. Recall that each of the nuisance parameters is given a separate degree of freedom in the northern and southern Galactic hemispheres.

In terms of these model parameters, we can then build up a likelihood function in terms of the binned data. In doing so, we treat each energy bin independently, so that in the $i^\text{th}$ bin we have:
\es{LLfunction}{
p_i\big(d_i \big| {\theta}_i\big)  = \prod_p { \mu^p_i(\theta_i)^{n^p_i} \,e^{-\mu^p_i(\theta_i)} \over n^p_i!} \,,
}
where $\mu^p_i(\theta_i)$ is the mean predicted number of photon counts in that pixel as a function of the model parameters $\theta_i = \{ {\bf \psi}_i, {\bf \lambda}_i \}$.  The $\mu^p_i(\theta_i)$ are calculated from the set of templates used in the fit, which describe the spatial distribution of the various contributions described above.  More specifically, if the $j^\text{th}$ template in energy bin $i$ predicts $T^{j,p}_i$ counts in the pixel $p$, then $\mu^p_i(\theta_i) = \sum_j A^j_i(\theta_i) \,T^{j,p}_i$, where $A^j_i(\theta_i)$ is the normalization of the $j^\text{th}$ template as a function of the model parameters.  In our analysis, all of the normalization functions are linear in the model parameters, and in particular there is a model parameter that simply rescales the normalization of each template in each energy bin.

The likelihood profile in the single energy bin, as a function of the parameters of interest $\psi_i$, is then given by maximizing the log likelihood over the nuisance parameters $\lambda_i$:
\es{LLprofile}{
\log  p_i\big(d_i \big| {\psi}_i\big) = \max_{\lambda_i} \log p_i\big(d_i \big| {\theta}_i\big) \,.
}
This choice to remove the nuisance parameters by taking their maximum is what defines the profile-likelihood method. After doing so we have reduced the likelihood to a function of just the DM parameters, which are equivalent to the isotropic and LOS integrated NFW correlated flux coming from DM decay. As such, we can write
\es{LLprofile2}{
\log  p_i\big(d_i \big| {\psi}_i\big) = \log p_i\Big(d_i \Big| \Big\{I^i_\text{iso}, I^i_\text{NFW}\Big\}\Big) \,.
}
For a given DM decay model, ${\cal M}$, there will be a certain set of values for $\{I^i_\text{iso}, I^i_\text{NFW}\}$ in each energy bin. Given these, the likelihood associated with that model is given by:
\es{LLprofile3}{
\log  p\big(d \big| {\cal M},  \{\tau, m_\chi\}\big) = \sum_{i=0}^{39}  \log p_i\Big(d_i \Big| \Big\{I^i_\text{iso}, I^i_\text{NFW}\Big\}\Big) \,,
}
where we have made explicit the fact that in most models the lifetime $\tau$ and mass $m_{\chi}$ are free parameters. We then define the test statistic~(TS) used to constrain the model ${\cal M}$ by\footnote{Note that this TS stands in contrast to that used in~\cite{Ackermann:2012rg}; in that work, the TS was similarly defined, except that instead of using $\tau = \infty$ as a reference the $\tau$ of maximal likelihood was used.  The definition of TS used here is more conservative than that in~\cite{Ackermann:2012rg}, though formally, with Wilk's theorem in mind, our limits do not have the interpretation of 95\% constraints.}
\es{TS}{
\text{TS}\big( {\cal M},  \{\tau, m_\chi \} \big) = 2 \times \Big[ \log  p\big(d \big| {\cal M},  \{\tau, m_\chi\}\big) - \log  p\big(d \big| {\cal M},  \{\tau = \infty, m_\chi\}\big)  \Big] \,.
}
Note that fundamentally it is the list of values $\{I^i_\text{iso}, I^i_\text{NFW}\}$ that determine the TS. This means we can build a 2-d table of TS values in each energy bin as a function of the extragalactic and Galactic DM flux. This table only needs to be computed once; afterwards a given model can be mapped onto a set of flux values, which has an associated TS in the tables. This is the approach we have followed, and we show these DM flux versus TS functions in Sec.~\ref{sec: detailed results}.  The table of TS values is also available as Supplementary Data~\cite{supp-data}.

\section{Likelihood Profiles}
\label{sec: detailed results}
\begin{figure}[b]
	\leavevmode
	\begin{center}$
	\begin{array}{c}
	\scalebox{0.33}{\includegraphics{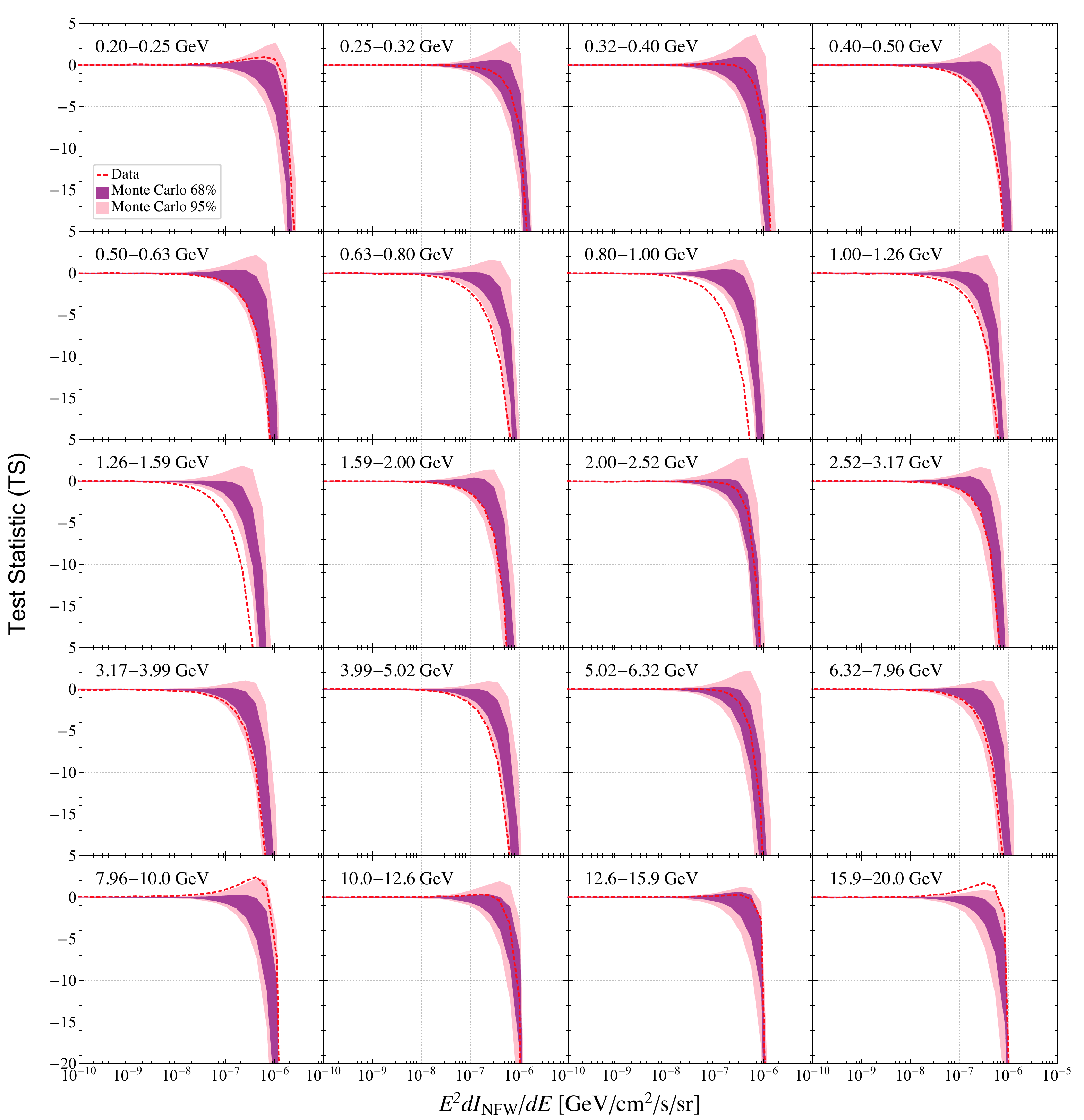}} 
	\end{array}$
	\end{center}
	\vspace{-.70cm}
	\caption{The change in log-likelihood, $\text{TS} \equiv p_i(d_i | \{ I_\text{NFW}^i \} ) -  p_i(d_i | \{ I_\text{NFW}^i = 0 \} )$, as a function of the intensity $I_\text{NFW}^i$ of NFW-correlated emission in the first 20 energy bins.  The measurement is given by the dashed red line, and the 68\% and 95\% confidence regions as derived from MC are given by the purple and pink bands respectively.  In most energy bins, the likelihood curves from the analysis of the data is seen to agree, within statistical uncertainties, with the expectation from the background templates only, as indicated by the MC bands. }
	\vspace{-0.15in}
	\label{Fig: LL_profileG0to19}
\end{figure}

\begin{figure}[t]
	\leavevmode
	\begin{center}$
	\begin{array}{c}
	\scalebox{0.33}{\includegraphics{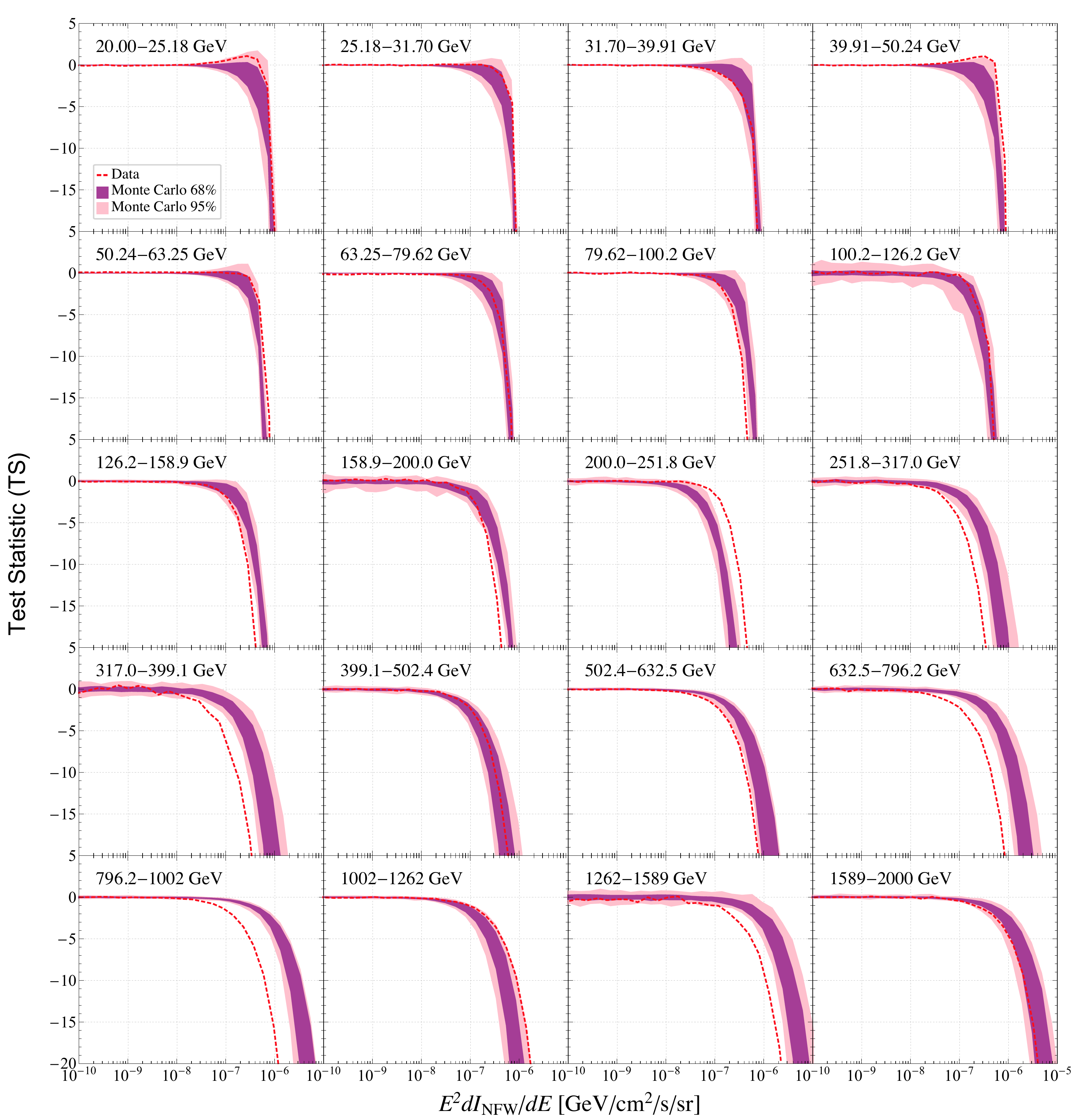}} 
	\end{array}$
	\end{center}
	\vspace{-.70cm}
	\caption{As in~\ref{Fig: LL_profileG0to19}, except for the later 20 energy bins.}
	\vspace{-0.15in}
	\label{Fig: LL_profileG20to39}
\end{figure}

As described in the main text, our limits on specific DM final states and models are obtained from 2-d likelihood profiles, where the two dimensions encompass LOS integrated NFW correlated Galactic gamma-ray flux and extragalactic gamma-ray flux.  In Figs.~\ref{Fig: LL_profileG0to19} and~\ref{Fig: LL_profileG20to39} we show slices of these log-likelihood profiles when the extragalactic DM-induced flux is set to zero.  The bands indicate the 68\% and 95\% confidence intervals for the expected profiles obtained from background-only MC simulations.  The simulations use the set of background (``nuisance") templates normalized to the best-fit values obtained from a template analysis of the data in the given energy bin.  In most energy bins, the results obtained on the real data are consistent with the MC expectations, showing that -- for the most part -- we are in a statistics-dominated regime.  In some energy bins, such as that from $15.9$--$20.0$ GeV, the data shows a small excess in the TS compared to the MC expectation.  While such an excess is perhaps not surprising since we are looking at multiple independent energy bins, it could also arise from a systematic discrepancy between the background templates and the real data.  More of a concern are energy bins where the limits set from the real data are more constraining than the MC expectation, such as the energy bin from $0.5$--$0.63$ GeV.  It is possible that this discrepancy, in part, arises from an over-subtraction of diffuse emission in certain regions of sky since the diffuse template is not a perfect match for the real cosmic-ray induced emission in our Galaxy.  This possibility -- and the efforts that we have taken to minimize its impact---is discussed further in Sec.~\ref{sec: systematics}.    

\begin{figure}[h]
	\leavevmode
	\begin{center}$
	\begin{array}{c}
	\scalebox{0.33}{\includegraphics{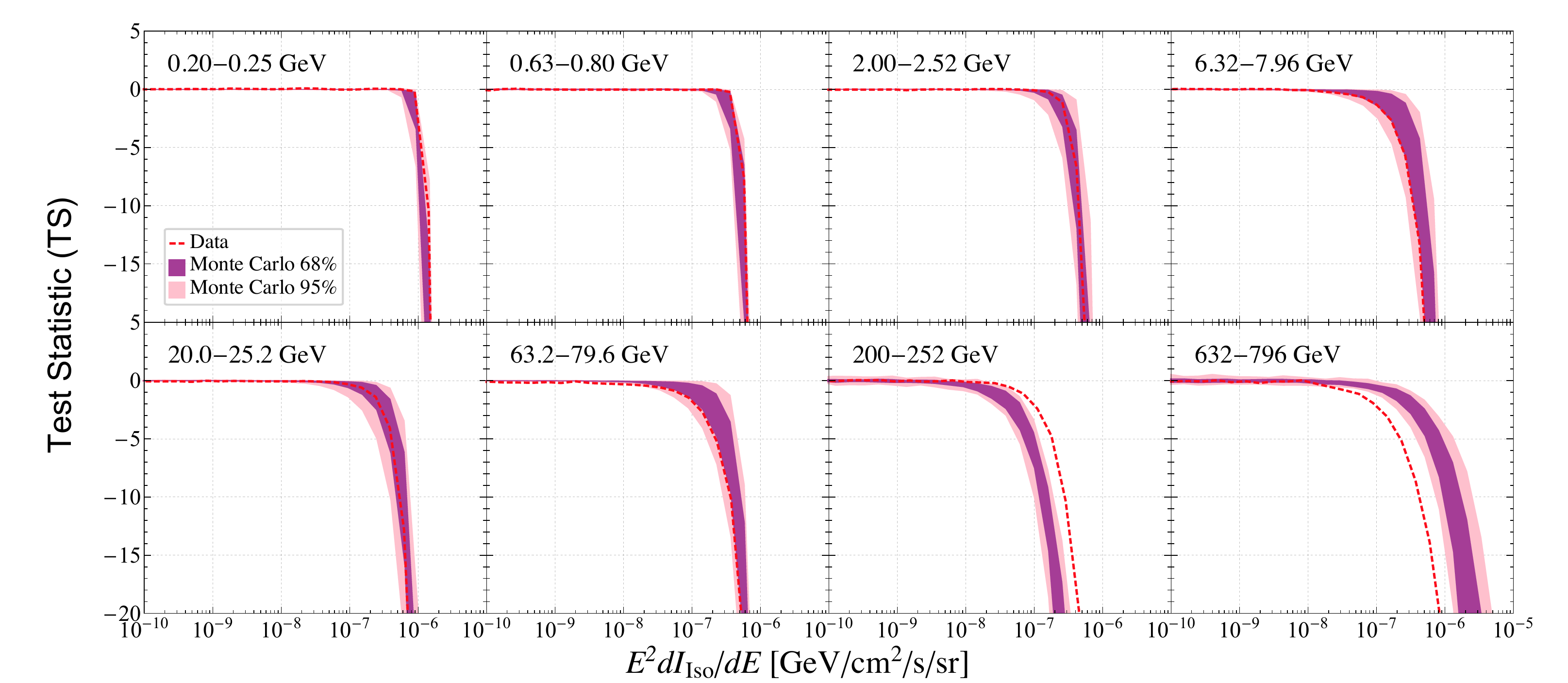}} 
	\end{array}$
	\end{center}
	\vspace{-.70cm}
	\caption{As in~\ref{Fig: LL_profileG0to19}, except for a selection of energy bins for the extragalactic only flux.}
	\vspace{-0.15in}
	\label{Fig: LL_profileEG}
\end{figure}

In Fig.~\ref{Fig: LL_profileEG}, we show a selection of the log-likelihood profiles found for vanishing Galactic DM-induced gamma-ray flux and shown instead as functions of the extragalactic DM-induced flux. It is important to remember that in the template fit we marginalize over isotropic emission.  As a result, it is impossible with our method to find a positive change in the TS as we increase the DM-induced isotropic flux $I_\text{iso}$.  In words, we remain completely agnostic towards the origin of the IGRB in our analysis.  That is, we do not assume the IGRB is due to standard astrophysical emission but we also do not assume it is due to DM decay. The 1-d likelihood profiles as functions of $I_\text{iso}$ instead show the limits obtained for the isotropic flux coming simply from the requirement that they do not overproduce the observed data.

In some energy bins, particularly at high energies (such as the energy bin from 632-796 GeV in Fig.~\ref{Fig: LL_profileEG}), the data is seen to be more constraining than the MC expectation.  However, we stress that the isotropic flux is not well determined, especially at these high energies, in our small region.  With that said, the isotropic flux determined in this small region tends to be larger than the IGRB determined from a dedicated analysis at high latitudes (see Fig.~\ref{Fig: DMvsBkgSpec}).  As a result, our limits on the extragalactic flux are likely conservative.

The full 2-d likelihood profiles are available as Supplementary Data~\cite{supp-data}. These are given as a function of the average Galactic and extragalactic DM flux in our ROI, without including any point source mask. The absence of the point source mask is chosen to simplify the use of our flux-TS tables.

\section{Systematics Tests}
\label{sec: systematics}

\begin{figure}[htb]
	\leavevmode
	\begin{center}$
	\begin{array}{cc}
	\scalebox{0.33}{\includegraphics{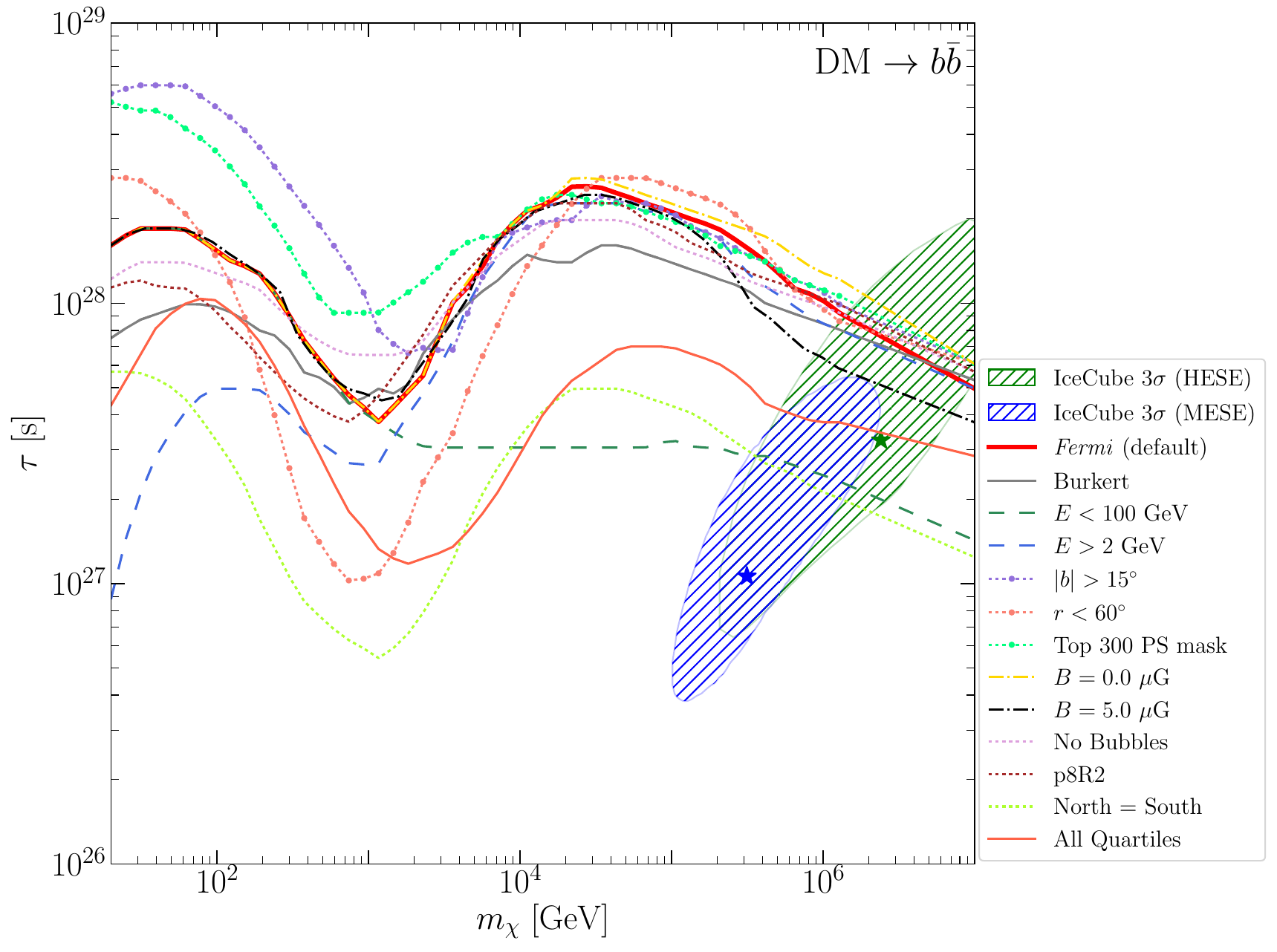}} & \scalebox{0.33}{\includegraphics{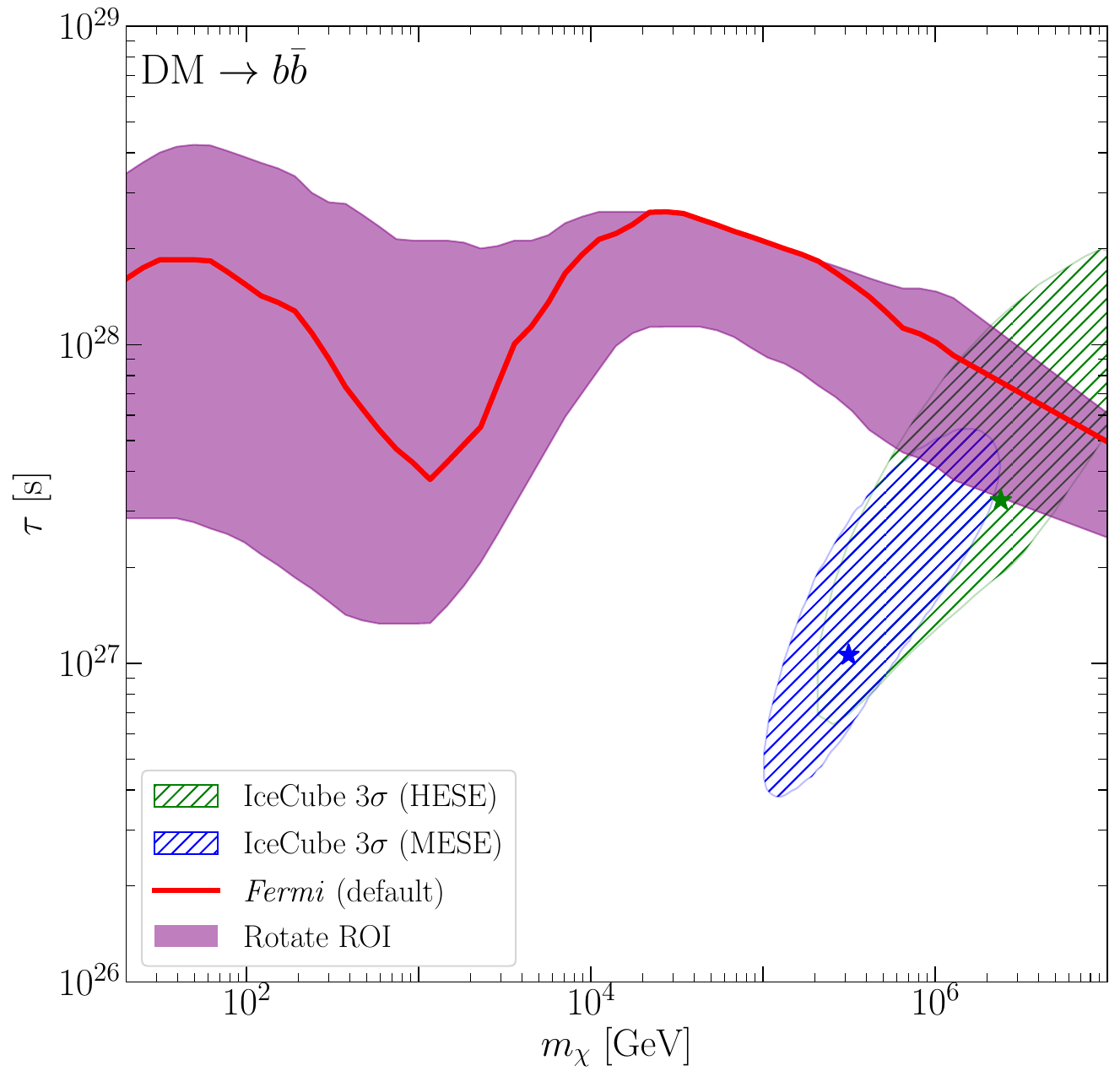}}
	\end{array}$
	\end{center}
	\vspace{-.70cm}
	\caption{Left: The limit derived for DM decay to $b \bar{b}$ for ten systematic variations on our analysis, as compared to our default analysis. Right: A purely data driven systematic cross check, where we have moved the position of our default ROI to five non-overlapping locations around the Galactic plane ($b=0$) and show the band of the limits derived from these regions is consistent with what we found for an ROI centered at the GC. See text for details.}
	\vspace{-0.15in}
	\label{Fig: DataSystematics}
\end{figure}

We have performed a variety of systematic tests to understand the robustness of our analysis.  Figure~\ref{Fig: DataSystematics} summarizes the results of some of the more important tests. 

In Fig.~\ref{Fig: DataSystematics}, we show limits on the $b\, \bar b$ final state with a variety of different variations on the analysis method.  Certain variations are shown to cause very little difference, such as not including an extra {\it Fermi} bubbles template, taking $B = 0.0$ $\mu$G when computing the IC flux, and using the more up-to-date Pass 8 model \texttt{gll\_iem\_v06} (\texttt{p8r2}) diffuse model instead of the \texttt{p7v6} model. As the \texttt{p8r2} model identifies regions of extended excess emission in the data and adds these back to the model, it is unclear if such a model would absorb a potential DM signal. Due to this concern, we used the \texttt{p7v6} model as our default in the main analyses.

Assuming $B = 5.0$ $\mu$G when computing the IC flux leads to slightly weaker constraints at higher masses due to the decrease in the IC contribution, as would be expected.  However, we emphasize that Faraday rotation measurements suggest that $B \leq 2.0$ $\mu$G across most of our ROI~\cite{2012ApJ...755...21M}, so $5.0$ $\mu$G is likely overly conservative.

We also note that the limit $B \to 0.0$ $\mu$G must be taken with care.  Without any magnetic field, the energy loss rate of high energy electrons and positrons from IC emission alone is not sufficient to keep the leptons confined to the halo.  However, even taking $B \sim 0.1-1$ nG, which is a typical value quoted for intergalactic magnetic fields, the Larmor radius $\sim$$0.1~(E_e /100 \, \, \text{TeV})(1\, \, \text{nG}/B)$ kpc, with $E_e$ the lepton energy, is sufficiently small to confine the $e^{\pm}$ in our ROI.  Larger values of circumgalactic magnetic fields in the halo are more likely.

An additional systematic is the assumption of the DM profile, as direct observations do not sufficiently constrain the profile over our ROI and we must rely on models.  In this work, we have assumed the NFW profile.  Another well-motivated profile is the Burkert profile~\cite{Burkert:1995yz}, which is similar to the NFW at large distances but has an inner core that results in less DM towards the center of the Galaxy.  In Fig.~\ref{Fig: DataSystematics} we show the limit we obtain using the Burkert profile with scale radius $r_0 = 13.33$ kpc.  From this analysis we conclude that the systematic uncertainty from the DM profile is less significant than other sources of uncertainty associated with the data analysis.

Masking the top 300 brightest and most variable PSs across the full sky, instead of masking all PSs, and masking the Galactic plane at $|b| > 15^\circ$, instead of $20^\circ$, both lead to stronger constraints at low energies. This is not surprising considering that the PS mask at low energies significantly reduces the ROI, and so any increase to the size of the ROI helps strengthen the limit.  Going out to distances within $60^\circ$ of the GC, on the other hand, slightly strengthens the limit at low masses, gives a similar limit at high masses, but weakens the limit at intermediate masses.  This is due to the fact that the diffuse templates often provide poor fits to the data when fit over too large of regions.  As a result, it becomes more probable that the added NFW-correlated template can provide an improved fit to the data, which is the case at a few intermediate energies.  This is also the reason why the limit is found to be slightly worse when the templates are not floated separately in the North and South, but rather floated together across the entire ROI (North=South in Fig.~\ref{Fig: DataSystematics}).  As a result, we find that the addition of the NFW-correlated template often slightly improves the overall fit to the data in this case.  Since it is hard to imagine a scenario where a DM signal would show up in the North=South fit and not in the fit where the North and South are floated independently, and since the latter analysis provides a better fit to the data, we float the templates independently above and below the plane in our main analysis.
Reassuringly, most of the systematics do not have significant effects at high masses, where we are generally in the statistics dominated regime.  

Many of the variations discussed above are associated with minimizing the impact of over-subtraction as discussed in the main text. Fundamentally, we do not possess a background model that describes the gamma-ray sky to the level of Poisson noise, and the choice of ROI can exacerbate the issues associated with having a poor background model. To determine our default ROI we considered a large number of possibilities and chose the one where we had the best agreement between data and MC, which ultimately led us to the relatively small ROI shown in Fig.~\ref{Fig:ROI} used for our default analysis. We emphasize that we did not choose the ROI where we obtained the strongest limits, as is clear from Fig.~\ref{Fig: DataSystematics}, and as such we do not need to impose a trials factor from considering many different limits.\footnote{We note that since this work appeared, an alternative physically motivated ``signal injection test" criteria for determining the analysis ROI has been promoted in the literature, see~\cite{Lisanti:2017qlb,Lisanti:2017qoz,Chang:2018bpt} for applications to DM searches using {\it Fermi} data.  The approach is to simply inject a hypothesis DM signal on top of either MC or the actual data.  One can then validate the analysis procedure by ensuring that the injected signal is not excluded. This procedure can yield more conservative choices than the requirement of data and MC agreement used here.  We do not expect that this modified criterion would move our bounds beyond the range spanned by Fig.~\ref{Fig: DataSystematics}, which demonstrates the impact of a variety of systematic effects.}

\begin{figure}[htb]
	\leavevmode
	\begin{center}$
	\begin{array}{cc}
	\scalebox{0.4}{\includegraphics{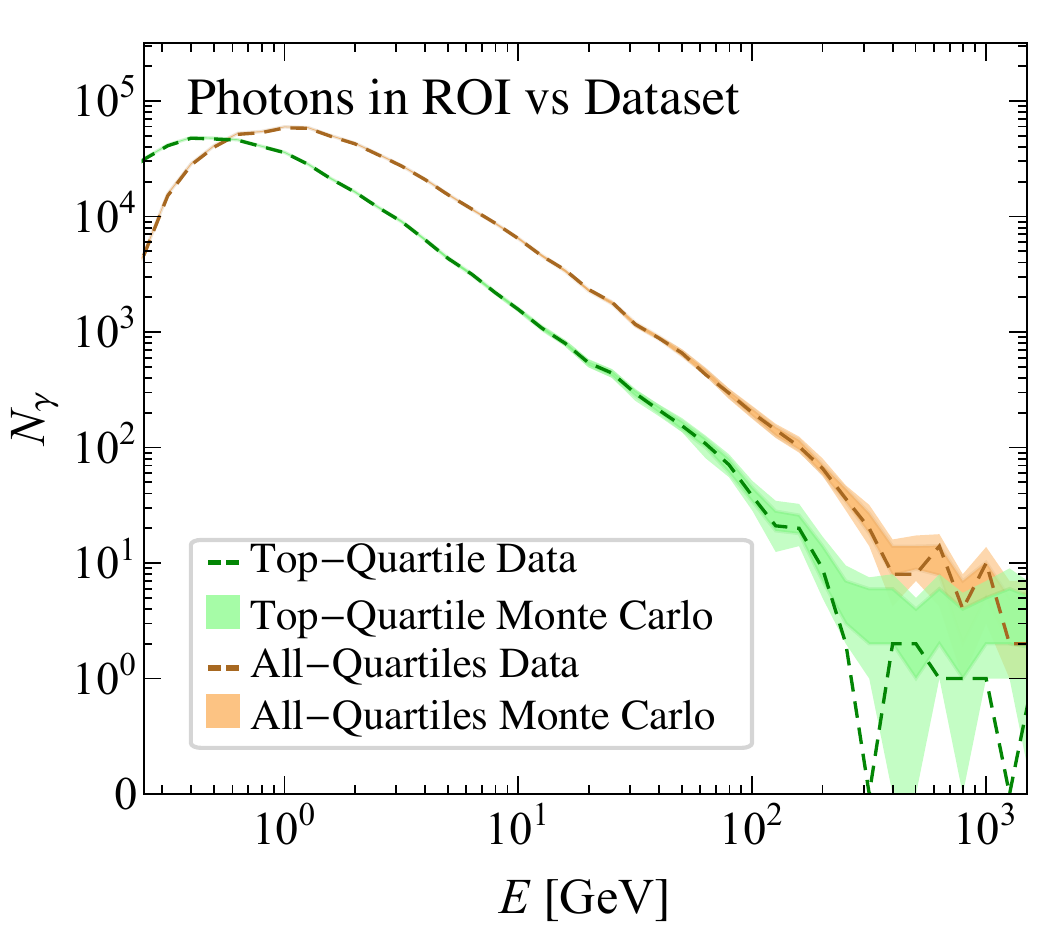}}  & \scalebox{0.4}{\includegraphics{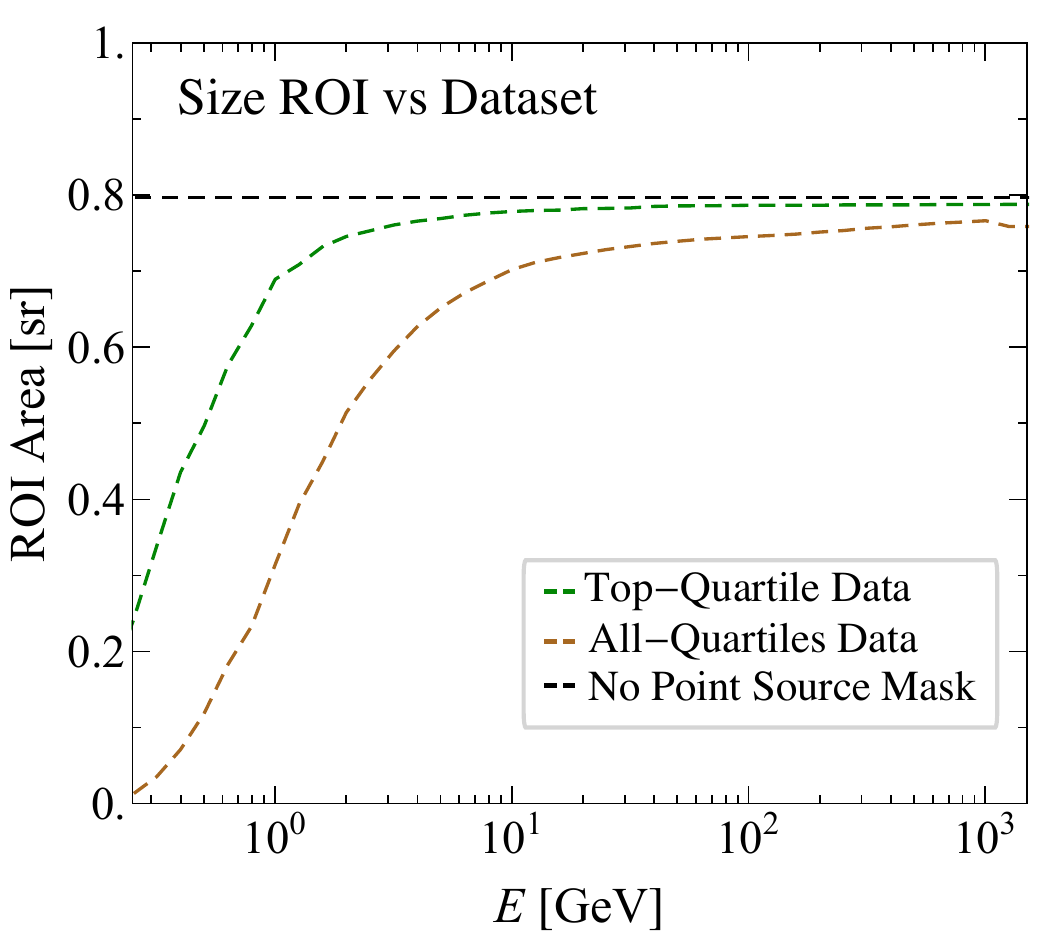}} \\
	\scalebox{0.4}{\includegraphics{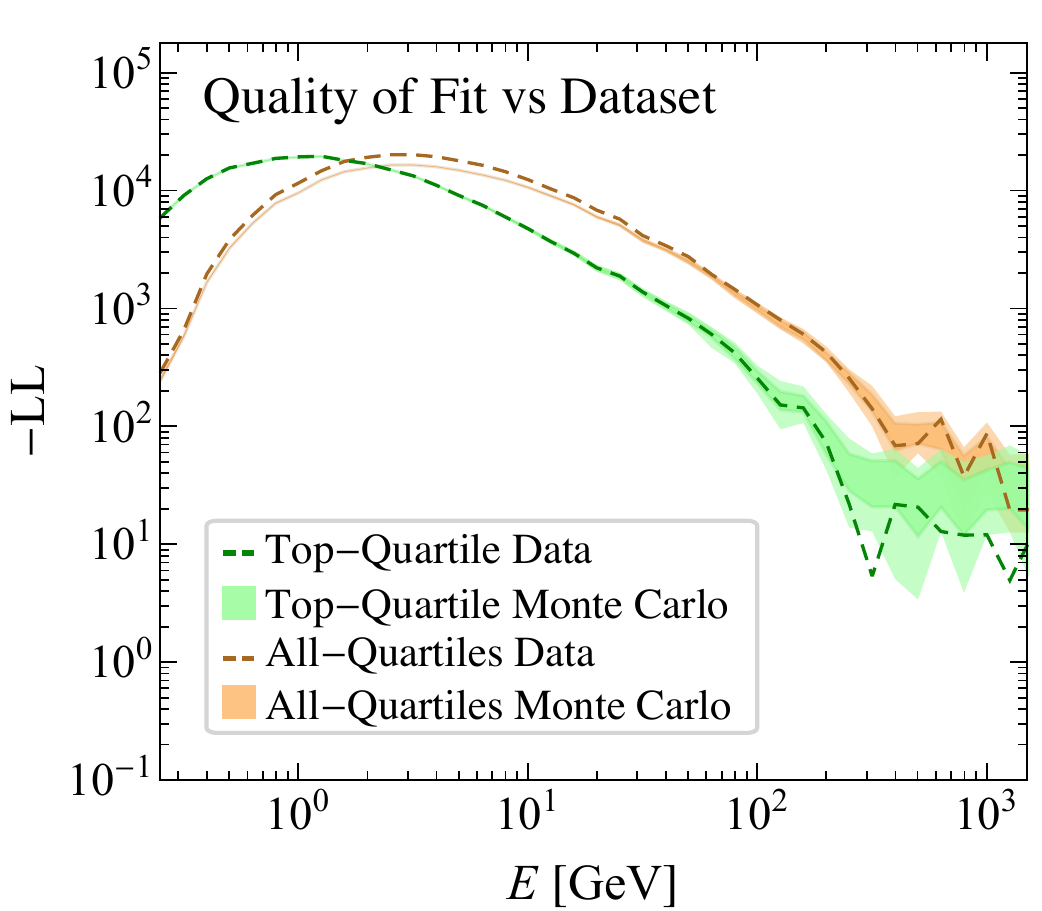}}  & \scalebox{0.41}{\includegraphics{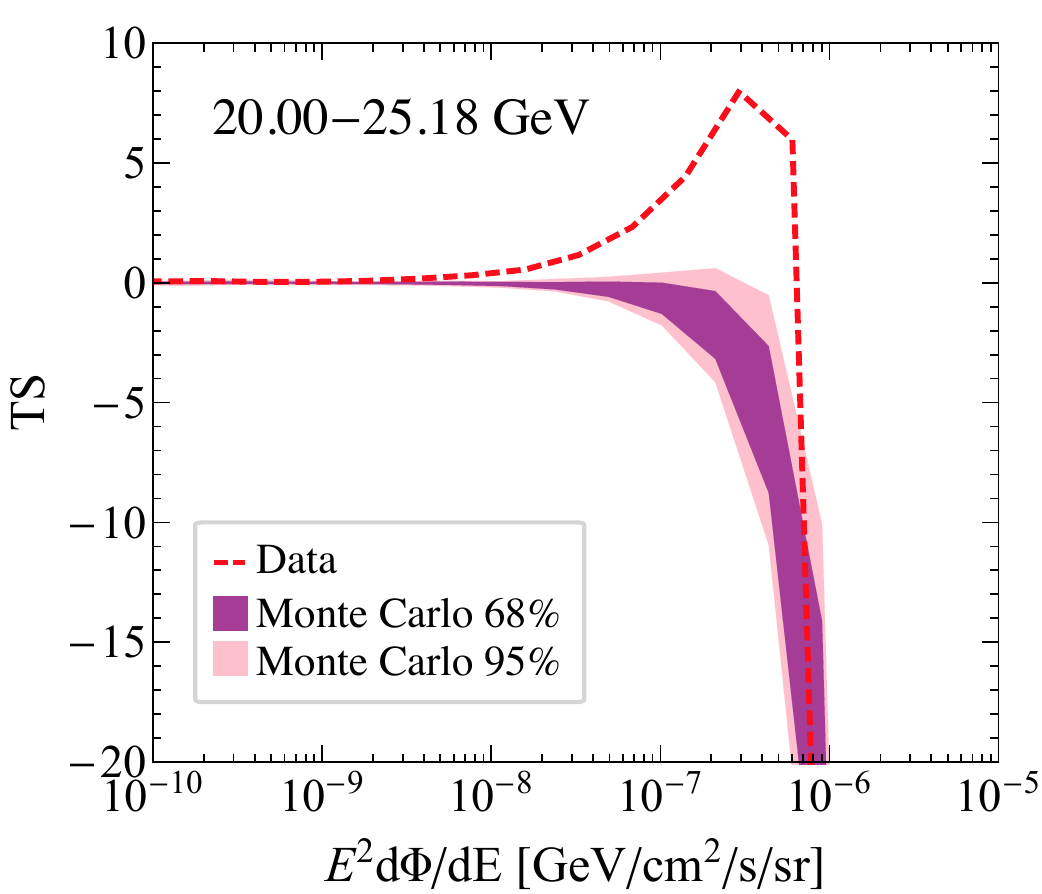}}
	\end{array}$
	\end{center}
	\vspace{-.70cm}
	\caption{Top left: the number of photons in our ROI as a function of energy for the top-quartile and all-quartiles. We show both the result in data and MC, where for the MC we indicate the 68\% and 95\% confidence intervals constructed from multiple MC realizations in each energy bin. Top right: the size of the ROI, in sr, as a function of energy. The variation with energy is due to the changing size of the PS mask. Bottom left: As in the top left plot, but here we show the quality of fit (the negative of the log-likelihood) as a function of energy. Bottom right: At intermediate energies there are residuals in the all-quartile data that can be absorbed by our Galactic DM template, leading to large excesses such as the one shown here. Such excesses play a role in the all-quartile limit being weaker than the top-quartile limit, as shown in Fig.~\ref{Fig: DataSystematics}.  However, the all-quartile limits are also weaker in part due to the reduced ROI at low energies.}
	\vspace{-0.15in}
	\label{Fig: UCVA}
\end{figure}

A further important systematic is our choice of data set.  In our main analysis, we used the top quartile of events, as ranked by the PSF, from the UltracleanVeto class.  Roughly four times as much data is available, within the same event class, if we take all photons regardless of their PSF ranking.  Naively using all of the available data would strengthen our bound.  However, as we show in Fig.~\ref{Fig: DataSystematics}, this is not the case---in fact, the limit we obtain using all photons is weaker than the limit we obtain using the top quartile of events.  There are two reasons for this, both of which are illustrated in Fig.~\ref{Fig: UCVA}. The first reason is simply that since we mask PSs at 95\% containment, as determined by the PSF, there is less area in our ROI in the analysis that uses all quartiles of events relative to the analysis using the top quartile.  Indeed, in Fig.~\ref{Fig: UCVA} we show the number of counts $N_\gamma$ in the different energy bins in our ROI for the top-quartile and all-quartile analyses.  At high energies, the top-quartile analysis has fewer photons than the all-quartile analysis, as would be expected.  However, since the PSF becomes increasingly broad at low energies, we find that at energies less than around 1 GeV, the top-quartile data has a larger $N_\gamma$.  Since both the IC and extragalactic emission tends to be quite soft, the data at low energies has an important impact on the limits. We further emphasize this by showing the size of the ROI as a function of energy in Fig.~\ref{Fig: UCVA}.
 
The second difference between the two data sets is that with the top-quartile only we find that the data is generally consistent with the background models, up to statistical uncertainties, while with the full data set there are systematic differences between the data and background models across almost all energies.  This is illustrated in the bottom left panel of Fig.~\ref{Fig: UCVA}, where we compare the data result to expectations from MC (68\% and 95\% statistical confidence intervals) from the background templates only for the maximum log-likelihood. There we see that in the top quartile case the data is consistent with MC up to energies $\sim$100 GeV.  In the all-quartile case, on the other hand, the data appears to systematically have a larger log-likelihood than the MC at energies less than around 50 GeV.  This difference could again be due to the increased PSF in the all quartile case, which smears out small errors in background mis-modeling onto larger scales. The addition of a Galactic DM template can then be used to improve this mis-modeling, which can lead to a strong preference for the DM decay flux in isolated energy bins, an example of which is shown in the bottom right panel of Fig.~\ref{Fig: UCVA}. Such excesses weaken the limit that can be set and ultimately play a central role in the all-quartile limit being weaker than naively expected.
 
We note that even in the top-quartile case there does appear to be some systematic difference between the MC expectation and the data at energies greater than around 100 GeV.  In particular, the data appears to generally have fewer photons than expected from MC.  With that said, this is a low-statistics regime where some energy bins have $N_\gamma = 0$.  This difference is also not too surprising, considering that the PS model and diffuse model were calibrated at lower energies and simply extrapolated to such high energies.  Part of this difference could also be due to cosmic ray contamination.  Thus, systematic discrepancies between data and MC at energies greater than around 100 GeV should be expected.  To illustrate the importance of this high energy data on our results, we show in Fig.~\ref{Fig: DataSystematics} the limit obtained when only including data with photon energies less than $100$ GeV; at 10 PeV, the limit is around 5 times weaker without the high-energy data. We also show in that plot the impact of removing the data below $2$ GeV, which has a large impact at lower masses but a minimal impact at higher masses.

In addition to the numerous variations of our modeling discussed above, we have also performed a purely data driven systematic cross check on our analysis shown on the right of Fig.~\ref{Fig: DataSystematics}, similar to that used in \cite{Calore:2014xka}. In the absence of any DM decay flux in the {\it Fermi} data, there should be nothing particularly special about the ROI near the Inner Galaxy that we have used---shown in Fig.~\ref{Fig:ROI}---and we should be able to set similar limits in other regions of the sky. This is exactly what we confirm in Fig.~\ref{Fig: DataSystematics}, where in addition to our default limit we show the band of limits derived from moving our ROI to five non-overlapping regions around the Galactic plane ($b=0$). As shown in the figure, even allowing for this data driven variation, the best fit IceCube points always remain in tension with the limit we would derive.

As a final note, we emphasize the importance of modeling non-DM contributions to the gamma-ray data in addition to the spatial morphology of the signal.  The limits on the DM lifetime would be weaker if we used a more simplistic analysis that did not incorporate background modeling and spatial dependence into the likelihood.  For example, we may set a conservative limit on the DM lifetime by using a likelihood function 
\es{LLsimp}{
\log p_i(d | \psi) = \sum_i \max_{\lambda_i}  \left[- {\left(  \sum_p \mu_i^p(\psi, \lambda_i) - \sum_p n_i^p \right)^2 \over 2  \sum_p \mu_i^p(\psi, \lambda_i)  }  - {1 \over 2} \log \left( 2 \pi  \sum_p \mu_i^p(\psi, \lambda_i) \right) \right]\,. 
}
The likelihood function depends on $ \sum_p \mu_i^p(\psi, \lambda_i) \equiv \sum_p \mu_i^p(\psi) +  \lambda_i$, which is a function of the  DM model parameters $\psi$.  The $ \lambda_i$ are nuisance parameters that allow us to add an arbitrary (positive) amount of emission in each energy bin.  These nuisance parameters account for the fact that we are assuming no knowledge of the mechanisms that would yield the gamma-rays recorded in this data set---the data may arise from DM decay or from something else.  As a corollary to this point, we may only determine limits with this likelihood function; by construction, we cannot find evidence for decaying DM.   Using~\eqref{LLsimp} within our ROI, we estimate a limit $\tau \approx 1 \times 10^{27}$ s for DM decay to $b \, \bar{b}$ with $m_\chi=1\,$PeV.  This should be contrasted with the limit $\tau \approx 1 \times 10^{28}$ s that we obtain with the full likelihood function, as given in Eq.~\eqref{LLfunction}.  This emphasizes the importance of including spatial dependence and background modeling in the likelihood analysis, as this knowledge increases the limit by around an order of magnitude in this example.  Even more important is the inclusion of energy dependence in~\eqref{LLsimp}.  Were we to modify~\eqref{LLsimp} to only use one large energy bin from $200$ MeV to $2$ TeV, then the limit would drop to $\sim$$10^{25}$ s in this example.  However, it is important to emphasize that the DM-induced flux is orders of magnitude larger than the data at high energies for this lifetime.

\section{Extended Theory Interpretation}
\label{sec: models}

In this section, we expand upon the decaying DM interpretation of our results in the context of additional final states and also specific simplified models.  We begin by giving limits on a variety of two-body final states.  Then, we comment on how we may use universal scaling relations to extend our result to high DM masses, beyond where it is possible to generate the spectra in \textsc{Pythia}.  Next, we discuss generalities related to constructing EFTs for decaying DM.  This allows us to emphasize how in the context of a consistent DM theory---one that respects the gauge symmetries of the SM---there are usually multiple correlated final states. Finally, we illustrate this point by providing two example models. The limits on all final states and models considered in this work are provided as part of the Supplementary Data~\cite{supp-data}.

\subsection{Additional Final States}
\label{sec: additional}
\begin{figure}[htb]
	\leavevmode
	\begin{center}$
	\begin{array}{ccc}
	\scalebox{0.4}{\includegraphics{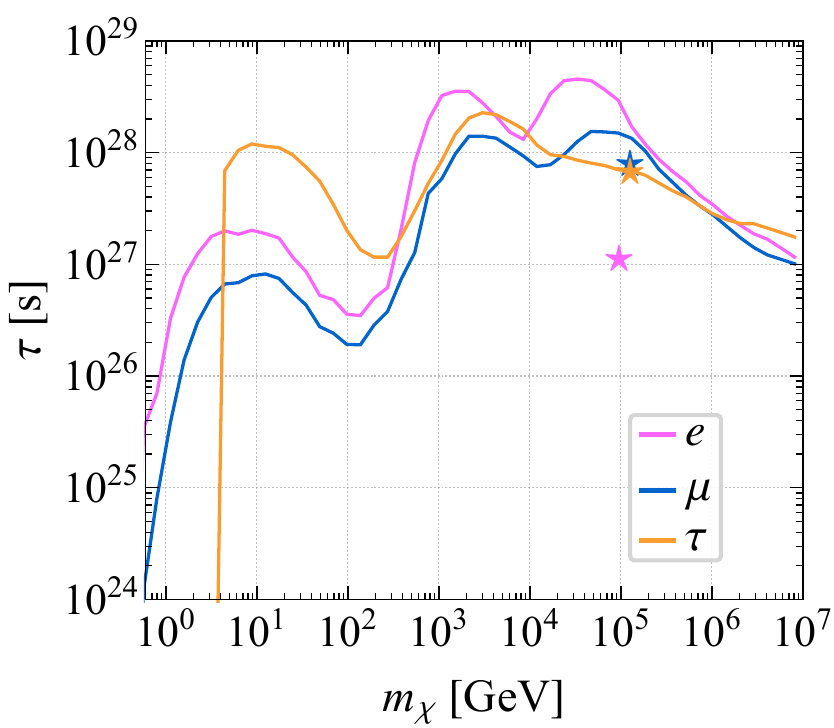}} & \scalebox{0.4}{\includegraphics{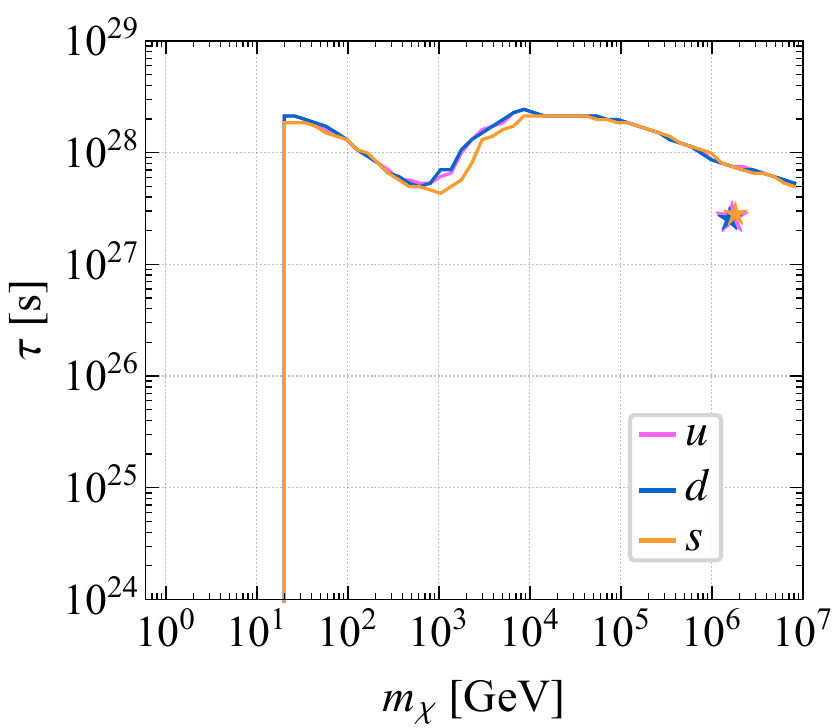}} & \scalebox{0.4}{\includegraphics{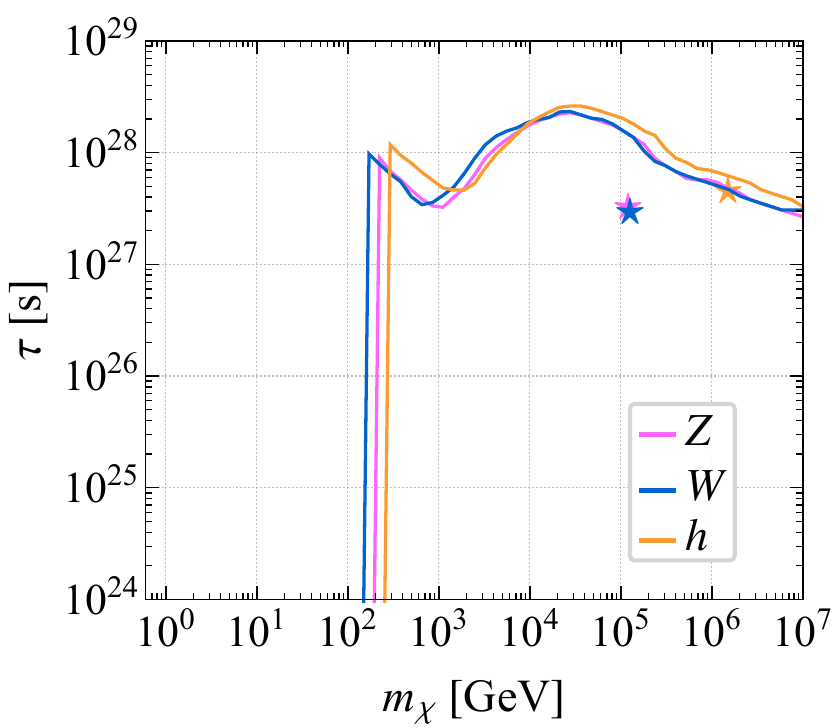}} \\
	\scalebox{0.4}{\includegraphics{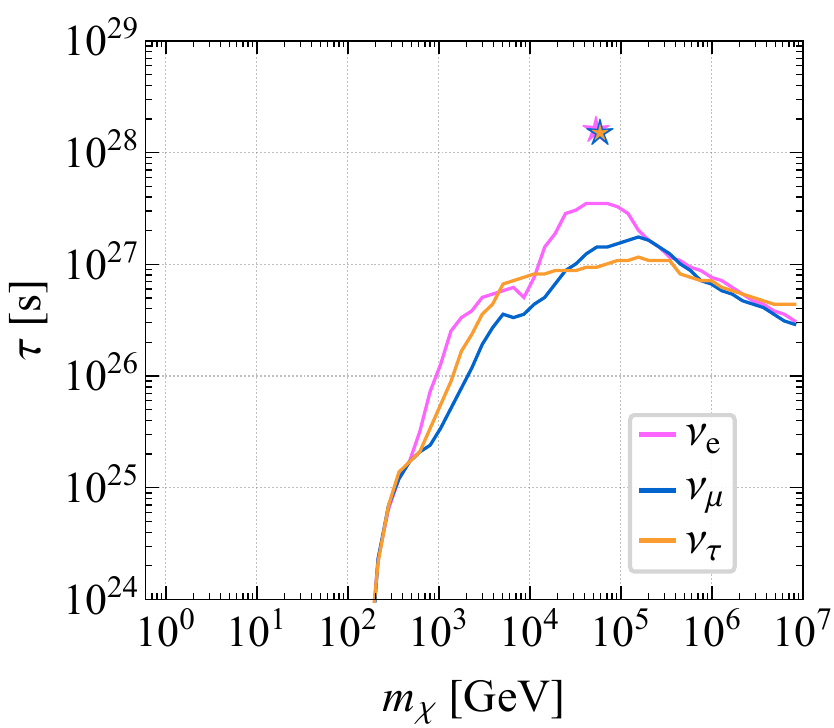}} & \scalebox{0.4}{\includegraphics{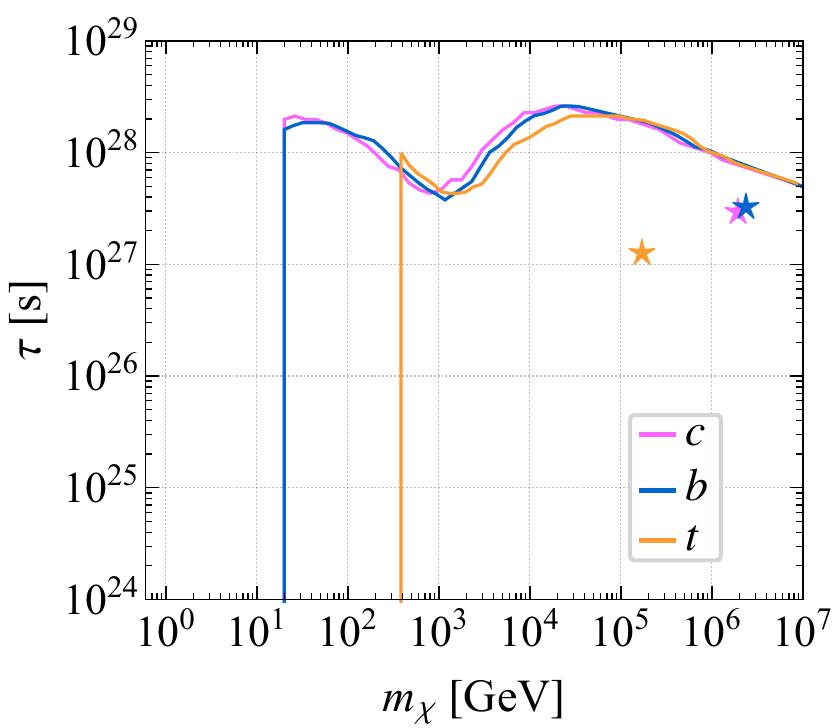}} & \scalebox{0.4}{\includegraphics{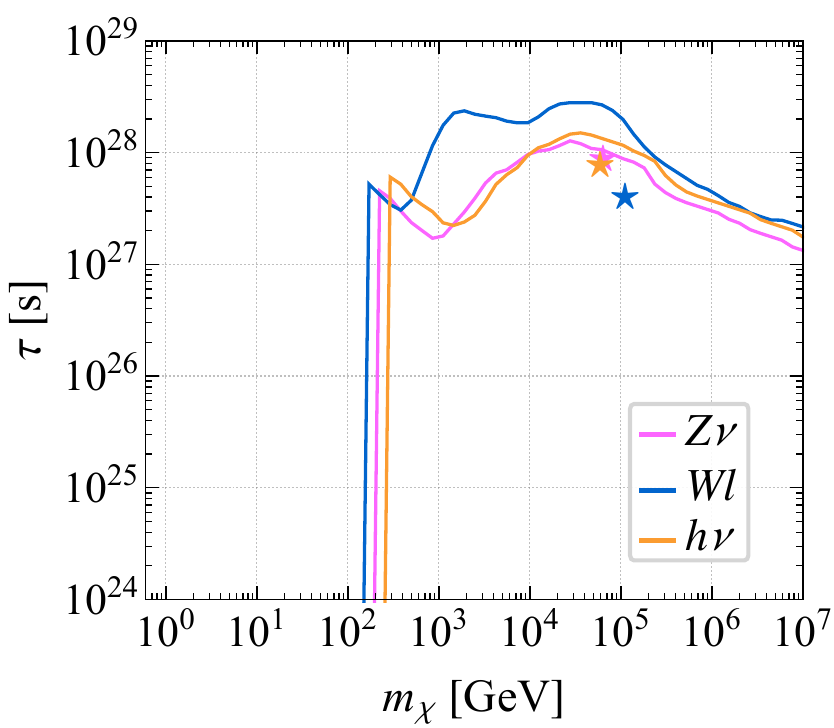}} \\ 
	\end{array}$
	\end{center}
	\vspace{-.70cm}
	\caption{Limits on all final states considered in this work. For each final state we show both the limit on the decay lifetime as a function of the DM mass, and also the best fit point for an interpretation of the IceCube flux with this channel as a star. Except for decay directly into neutrinos, for every other final state this best fit point is in tension with the limit we derive from {\it Fermi}.}
	\vspace{-0.15in}
	\label{Fig:OtherFSLims}
\end{figure}

In addition to DM decays directly to bottom quarks, the benchmark final state used extensively in this Letter, we also determine the {\it Fermi} limits on DM decay into a number of two-body final states. In detail we consider all flavor conserving decays to charged leptons, neutrinos, quarks, electroweak bosons, and Higgs bosons.  Due to our emphasis on modes that yield high energy neutrinos, we also include three mixed final states, $Z \nu$, $W \ell$ and $h \nu$. For these last three cases we consider an equal admixture of lepton and neutrino flavors. These limits are all shown in Fig.~\ref{Fig:OtherFSLims}. 

\begin{figure}[htb]
	\leavevmode
	\begin{center}$
	\begin{array}{c}
	\scalebox{0.37}{\includegraphics{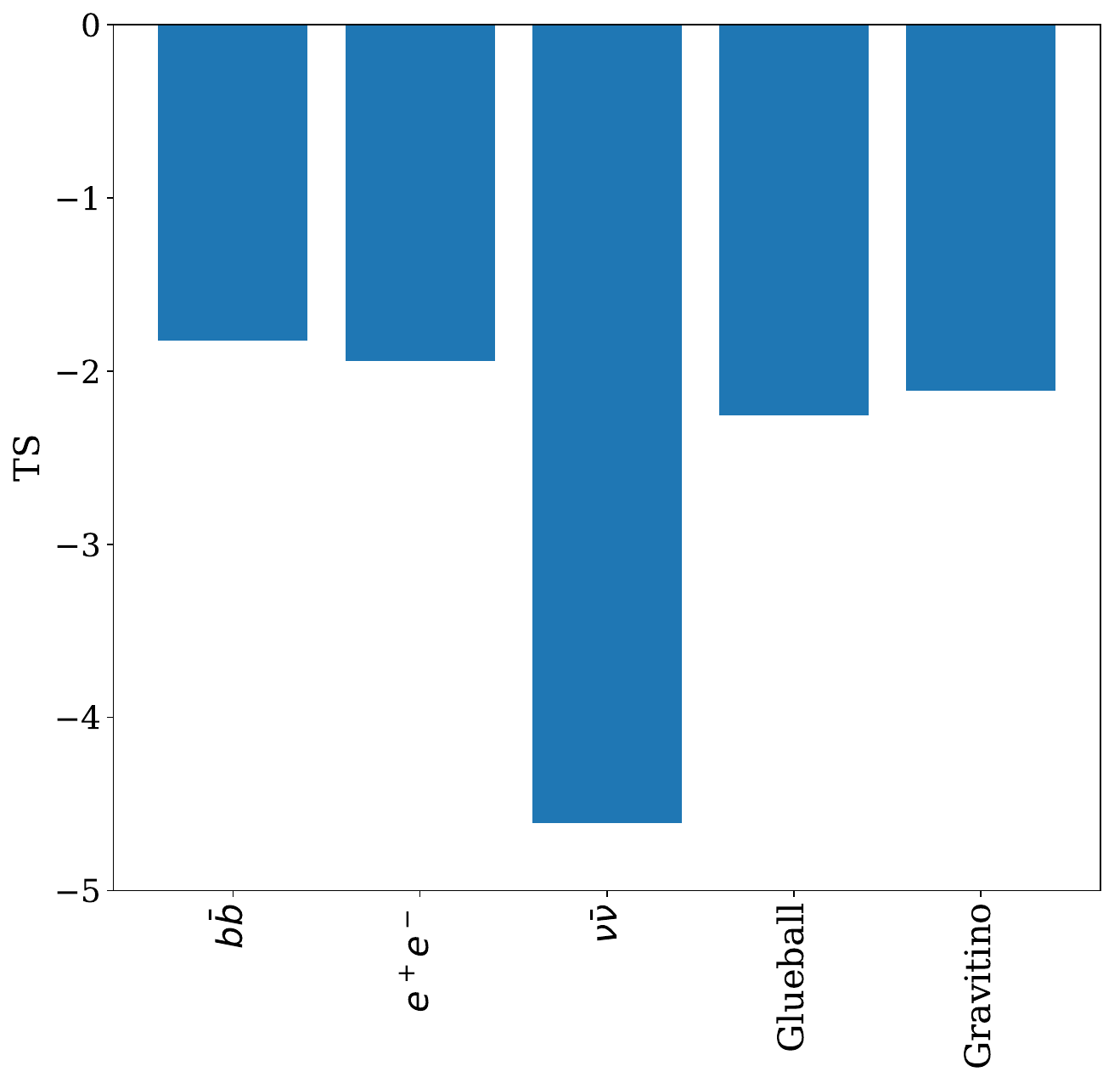}}
	\end{array}$
	\end{center}
	\vspace{-.70cm}
	\caption{Quality of the best fit to the combined IceCube data for a selection of final states and models. The quality of the fit is represented as a TS for the DM-only model defined with respect to the best fit power law with an exponential cutoff; this simple model is meant to represent an astrophysical fit to the data. Among the DM-only models, $b\, \bar b$ provides the best fit to the data, which motivated our choice to focus on it in the main body. Other final states and models give a comparative goodness of fit, except for the case of decay directly to neutrinos which gives a sharp spectrum and consequently a poor fit.}
	\vspace{-0.15in}
	\label{Fig:bestFit}
\end{figure}

Figure~\ref{Fig:OtherFSLims} has some interesting features. Channels which produce more electrons and positrons tend to have stronger limits at high masses due to the associated Galactic IC flux. This is clear for DM decays to $e^+ e^-$ and also to $\nu_e\, \bar{\nu}_e$. Most of the quark final states lead to nearly identical limits; these channels produce a large number of pions regardless of flavor yielding a similar final state spectrum. The only difference is for the top quark, which first decays to $b W$, thereby generating a prompt spectrum which differs from the lighter quarks. Note that for the lighter quarks, the threshold is still always set at 20 GeV; \textsc{Pythia} does not operate below this energy since they do not simulate the full spectrum of QCD resonances.  We leave the extension of our results to lower masses for colored final states to future work.

In addition to the limits, we also show the best fit point for a fit to the IceCube data as a star for each channel.  The best fit point is always in tension with the limits we derive from {\it Fermi}, except for decays directly into neutrinos.  However, as we show in the next subsection, when modeling the DM interactions in a consistent theory context, one must rely on a very restricted setup to manifest exclusive decays into neutrino pairs.

The quality of fit for the different stars represented in Fig.~\ref{Fig:OtherFSLims} are not identical. This point is highlighted in Fig.~\ref{Fig:bestFit} where we show the quality of fit (for DM only) for three two-body final states, $b$, $e$, and $\nu_e$, as well as two models, glueball and gravitino dark matter. The quality of fit is shown with respect to the best fit power law multiplied by an exponential cutoff, chosen to represent an astrophysical fit to the data. The astrophysical model always gives the best fit, with the $b\, \bar{b}$ DM-only model a worse fit to the data by a TS $\sim 1.9$. A number of the other final states and models also give a comparable quality of fit to the $b \, \bar b$ final state, as their neutrino spectra are all broad enough to fit the data in a number of energy bins. This is not the case for $\nu_e\, \bar{\nu}_e$---the only final state not in tension with our limits from \textit{Fermi}---where the sharp neutrino spectrum can at most meaningfully contribute to a single energy bin. 

\subsection{Extending \textbf{\textit{Fermi}} limits beyond 10 PeV}
\label{sec:ExtendingToGUTScale}

\begin{figure}[t]
	\leavevmode
	\begin{center}$
	\begin{array}{cc}
	\scalebox{0.39}{\includegraphics{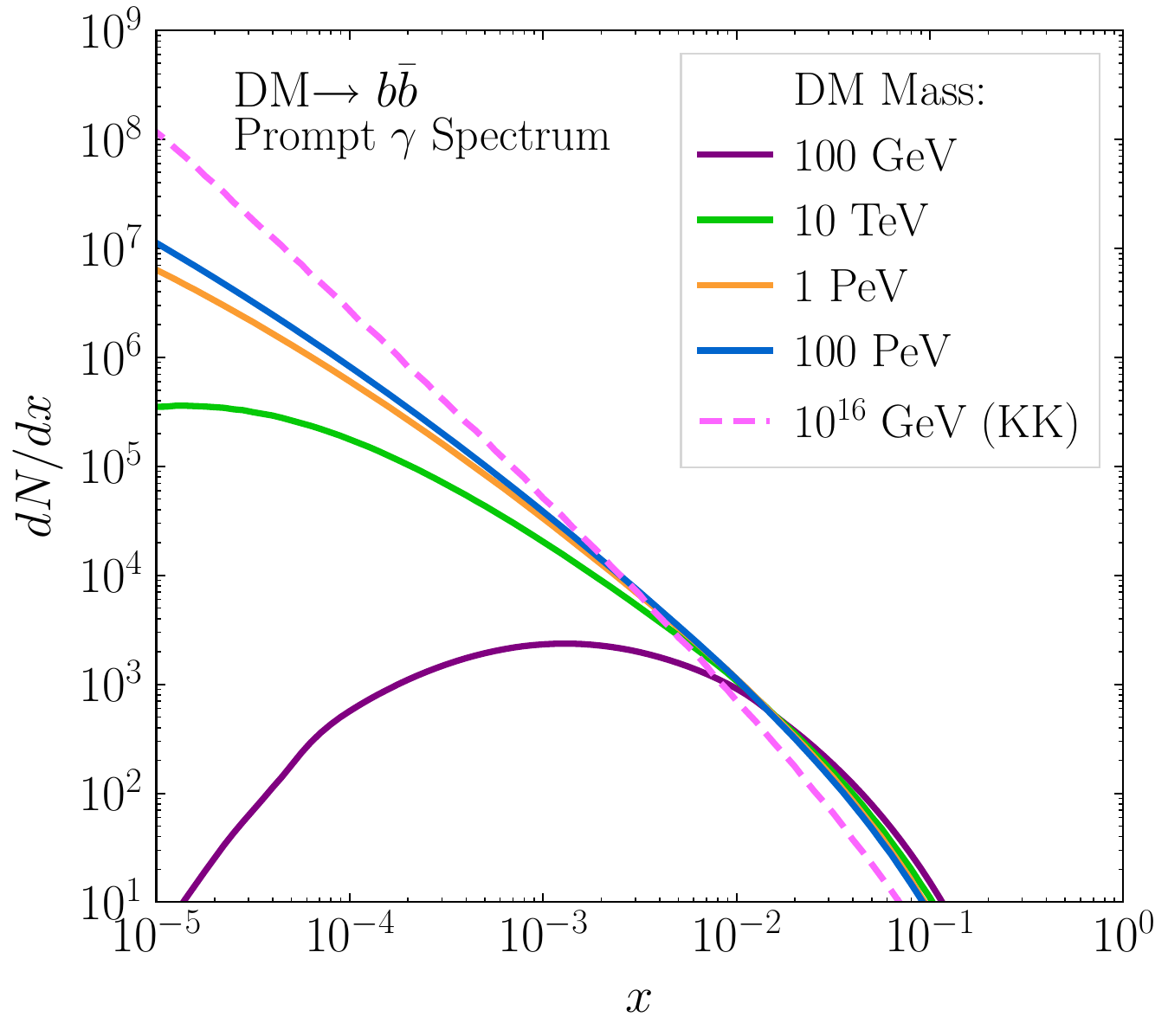}}  &
	\scalebox{0.51}{\includegraphics{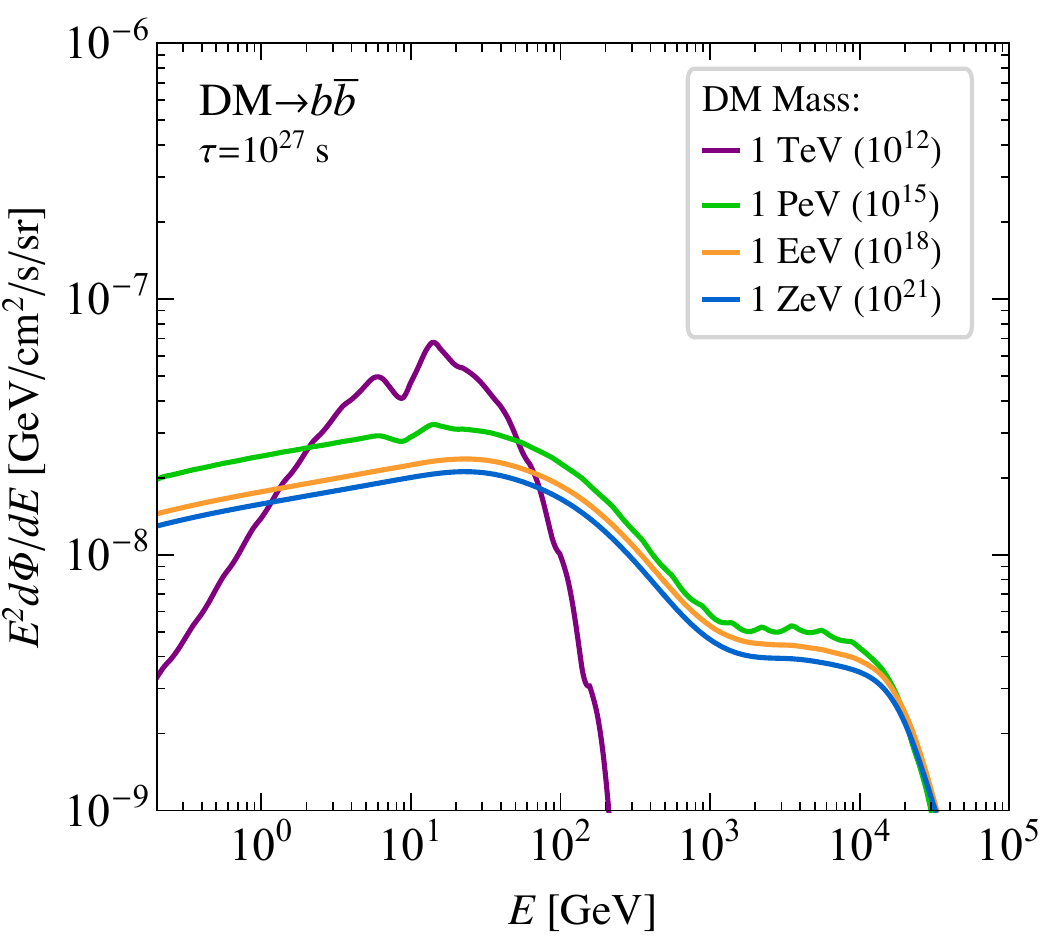}}
	\end{array}$
	\end{center}
	\vspace{-.70cm}
	\caption{Left: The prompt photon DM decay $b \,\bar{b}$ spectrum approaches a universal form in $dN/dx$, where $x = 2E/m$. All spectra except for the one at $10^{16}$ GeV were determined using \textsc{Pythia}; the spectrum at the GUT scale is taken from \cite{Kalashev:2016cre} and labelled as KK. The prompt $e^{\pm}$ and neutrino spectra also approach universal forms. Taken together this allows us to determine the $b \bar{b}$ spectrum up to masses $\sim$$10^{12}$ GeV. Right: At very high masses the Galactic flux from DM decay expected in the {\it Fermi} energy range is negligible. Nevertheless due to cascade processes, the extragalactic flux, shown here here for DM with $\tau = 10^{27}$ s, approaches a universal form. This implies {\it Fermi} can set an essentially mass independent limit on very heavy dark matter, as shown in Fig.~\ref{Fig:GUTLim}.}
	\vspace{-0.15in}
	\label{Fig:UniversalSpec}
\end{figure}
As discussed above, generating the prompt spectra much above 10\,PeV in \textsc{Pythia} is not feasible. The issue is already clear in Tab.~\ref{table:Particles}: as the DM mass is increased, so is the energy injected into the final state decays which leads to a large number of final states resulting from the showering and hadronization processes. At some point this process simply takes too long to generate directly. Nevertheless, this section provides the details of the spectrum generation for $b\, \bar{b}$, and then how these are utilized to extend our {\it Fermi} limits up to energies $\sim$$10^{12}$ GeV.

The key observation is that when the prompt photon, electron/positron, or neutrino spectra are considered in terms of $dN/dx$ where $x=2E/m_\chi$, for $b\, \bar{b}$, and likely many other channels though we have not fully characterized this for all final states, they approach a universal form independent of mass. This is shown for the case of photons on the left of Fig.~\ref{Fig:UniversalSpec}. There we show \textsc{Pythia} generated spectra up to 100 PeV, and compare them to a spectrum at the GUT scale determined in \cite{Kalashev:2016cre}. The computation in~\cite{Kalashev:2016cre} takes the fragmentation function for bottom quarks at lower energies, and then runs them to the GUT scale using the DGLAP evolution equations. This universality allows us to determine the prompt spectra for ${\rm DM} \to b\, \bar{b}$ with $m_\chi$ well above the PeV scale.

Given these spectra, the next consideration is whether a meaningful flux from these decays populates the {\it Fermi} energy range. For prompt and IC flux from the Milky Way the answer is no, as is evident already in Fig.~\ref{Fig: highlowbSpec}. The synchrotron flux from electrons and positrons is expected in the Fermi range or even higher energies for DM mass of $\gtrsim{10}^9$~GeV, which can improve the lifetime limits by a factor of 2-3~\cite{Murase:2012xs}. However, the results depend on halo magnetic fields that are uncertain. Thus, we here consider conservative constraints without the Galactic synchrotron component.  Nevertheless the situation is different for the extragalactic flux, as shown on the right of Fig.~\ref{Fig:UniversalSpec}. There we see that the amount of flux approaches a universal form, essentially independent of the DM mass. The intuition for how this is possible is as follows. The total DM energy injected in decays is independent of mass: as we increase the mass of each DM particle, the number density decreases as $1/m_{\chi}$, but at the same time the power injected per decay increases as $m_{\chi}$, keeping the total injected power constant. Extragalactic cascades reprocess this power into the universal spectrum shown, and this implies that above a certain mass the extragalactic flux seen in the {\it Fermi} energy range becomes a constant.

\begin{figure}[t]
	\leavevmode
	\begin{center}$
	\begin{array}{c}
	\scalebox{0.45}{\includegraphics{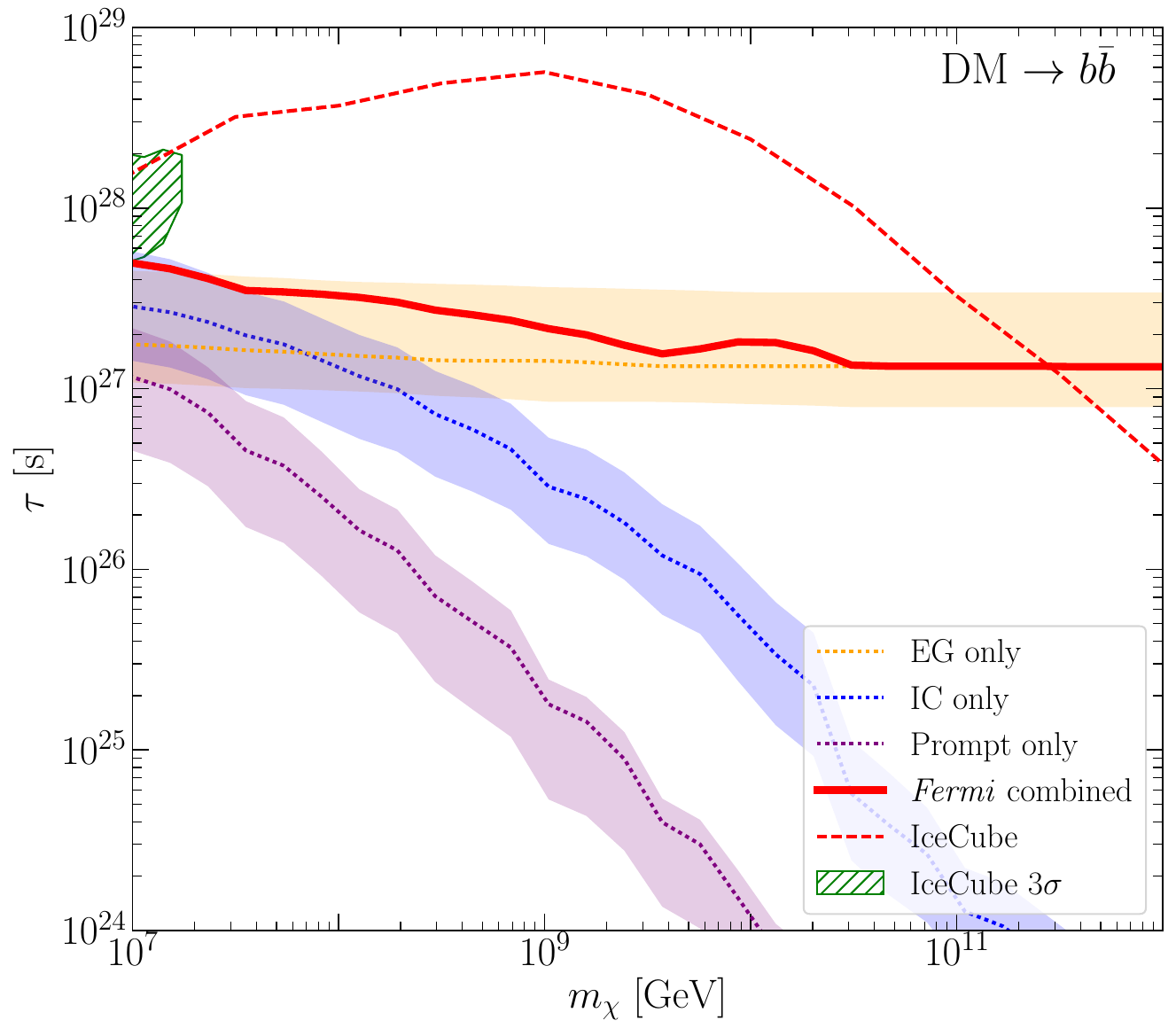}}
	\end{array}$
	\end{center}
	\vspace{-.70cm}
	\caption{Using the universal form of the spectra shown in Fig.~\ref{Fig:UniversalSpec}, {\it Fermi} can set limits on DM decays up to masses well above the PeV scale. At higher masses this limit comes only from the extragalactic contribution, such that after about $10^{10}$ GeV, the limit set becomes essentially independent of mass. Note that at these high masses, the {\it Fermi} limits are noticeably weaker than those obtained by direct searches for prompt Galactic photons from the decay of these heavy particles, as determined in \cite{Kalashev:2016cre}.  Note that the labeling is the same as in Fig.~\ref{Fig: glue} of the main Letter.}
	\vspace{-0.15in}
	\label{Fig:GUTLim}
\end{figure}

Using this, we extend our limits on the $b\, \bar{b}$ final state up to the masses $\sim$$10^{12}$ GeV in Fig.~\ref{Fig:GUTLim}. There we see that above $\sim$$10^{10}$ GeV, the limit becomes independent of mass and is coming only from the extragalactic contribution, exactly because of the universal form of the extragalactic flux. The same is not true for the neutrino spectrum---there is no significant reprocessing of the Galactic or extragalactic neutrino flux---and as such the limits IceCube would be able to set decrease with increasing mass. Despite this, limits determined from direct searches for prompt Galactic photons, which at these high energies are not significantly attenuated, set considerably stronger limits as shown in~\cite{Kalashev:2016cre}. Nevertheless, given that {\it Fermi} cannot see photons much above 2 TeV, we find it impressive that the instrument can set limits up to these masses.

We have cut Fig.~\ref{Fig:GUTLim} off at masses $\sim$$10^{12}$ GeV because at higher energies processes such as double pair-production may become important (see~\cite{Bhattacharjee:1998qc} for a review and references therein).  The neutrino limits may also be affected by scattering off the cosmic neutrino background at very high masses.  We leave such discussions to future work.

\subsection{EFT of Decaying Dark Matter}
\label{sec: EFT}

\begin{table}
\renewcommand{\arraystretch}{2.2}
\setlength{\arrayrulewidth}{.3mm}
\centering
\setlength{\tabcolsep}{0.4em}
\footnotesize
\begin{tabular}{|c|c|c|c|c|}
\hline
 $\Big(R_{SU(2)}\Big)_Y$ & operator & final states & ratios of  BR's, $m_\chi\gg\,$TeV  & $\tau\gtrsim10^{27}$ [s] \\[2pt]
\hline\hline
 \multicolumn{5}{|c|}{spin 0}\\
 \hline
\multirow{9}{*}{$(0)_0 $}  
& 	$\chi H^\dagger H $  & $hh$, $Z^0 Z^0$,$W^+ W^-$,$f\bar{f}$  & $1:1:2:16N_c y_f^2 \frac{v^2}{m^2_\chi}$ & $\bar{m}_\chi/\bar{\Lambda}^2\gtrsim 9 \times 10^{79}$\footnote{This operator corresponds to the glueball model. However, in that model the coefficient is naturally suppressed by dimensional transmutation.} \\ 
& 	\multirow{3}{*}{\vspace{14pt}$\chi\left( L H \right)^2$} & 
$\nu\nu hh$, 
$\nu\nu Z^0 Z^0$, 
$\nu\nu Z^0 h$, 
& $1:1:2:$
   & \multirow{3}{*}{$\bar{\Lambda}^4/\bar{m}_\chi^5\gtrsim 1$}  \\[-7pt]
& & $\nu e^- h W^{+}$, $\nu e^- Z^0 W^{+}$, $e^- e^- W^{+} W^{+}$, &
$2:2:4:$ & \\[-7pt]

&  & $\nu\nu h$, $\nu\nu Z^0$, $\nu e^- W^+$, $\nu\nu$ & 
$24 \pi^2 \frac{v^2}{m^2_\chi}\Big(1:1:1:768 \pi^2 \frac{v^2}{m^2_\chi}\Big)$ & \\
& 	$\chi H \bar{L}E$ &  $h\ell^+\ell^-$, $Z^0\ell^+\ell^-$, $W^\pm \ell^\mp \nu$, $\ell^+\ell^-$  & $1:1:2:32\pi^2 \frac{v^2}{m^2_\chi}$  &  $\bar{\Lambda}^2/\bar{m}_\chi^3\gtrsim 4\times 10^{29}$\\
& 	$\chi \tilde{H} \bar{Q} U$, $\phi H \bar{Q} D$  &  $h q \bar{q}$, $Z^0 q \bar{q}$, $W^\pm q' \bar{q}$, $q\bar{q}$ & $1:1:2:32\pi^2 \frac{v^2}{m^2_\chi}$ & $\bar{\Lambda}^2/\bar{m}_\chi^3\gtrsim 1\times 10^{30}$ \\
&	$\chi B_{\mu\nu} \ptwiddle{B}\,^{\mu\nu}$ & $\gamma\gamma$, $\gamma Z$, $ZZ$ & $c^4_W:2c^2_W s^2_W:s^4_W$ & $\bar{\Lambda}^2/\bar{m}_\chi^3\gtrsim 2\times 10^{31}$\\
&	$\chi W_{\mu\nu} \ptwiddle{W}\,^{\mu\nu}$ & $\gamma\gamma$, $\gamma Z^0$, $Z^0 Z^0$, $W^+ W^-$ \footnote{Additional three- and four-body decays are suppressed.} & $s^4_W:2c^2_W s^2_W:c^4_W:2$ &  $\bar{\Lambda}^2/\bar{m}_\chi^3\gtrsim 6\times 10^{31}$ \\
&	$\chi G_{\mu\nu} \ptwiddle{G}\,^{\mu\nu}$ & hadrons  & $1$ & $\bar{\Lambda}^2/\bar{m}_\chi^3\gtrsim 2\times 10^{32}$\\
&	$\chi D_\mu H^\dagger D^\mu H$ & $hh$, $Z^0 Z^0$, $W^+ W^-$ \footnote{$Z^0 Z^0 hh$ is further suppressed by  four-body phase space factors.} & $1:1:2$ & $\bar{\Lambda}^2/\bar{m}_\chi^3\gtrsim 3\times 10^{30}$ \\
\hline
\multirow{ 4}{*}{$(2)_{1/2}$\footnote{Here we are assuming that $\chi$ is a scalar.  The pseudo-scalar case can be inferred by making the appropriate replacements to conserve CP.  See the text for details.}}  
& 	$V_{\hat\lambda}$~\cite{Gunion:2002zf}\footnote{For brevity, we follow the notation of~\cite{Gunion:2002zf}, which studies the Two-Higgs-Doublet model in the decoupling limit.  $V_X$ denotes that the potential $V$ which governs the interactions between the heavy state and the SM is dominantly controlled by the coupling $X$.  See text for details} & $hhh$, $hZ^0 Z^0$, $hW^+ W^-$  & $1:1:2$ & $g^2 \bar{m}_\chi \lesssim 2 \times 10^{-53}$ \\
& 	$V_{c_{\beta-\alpha}}$~\cite{Gunion:2002zf}$^{e,}$\footnote{The mixing factor $c_{\beta-\alpha}\to v^2/m^2_\chi$ in the decoupling limit.} & $hh$, $Z^0 Z^0 $, $W^+ W^-$  & $\big(1+(\lambda_T-2\lambda_A)/\lambda\big)^2:1:2$ & $\bar{m}_\chi/c^2_{\beta-\alpha} \gtrsim 4 \times 10^{48}$ \\
& 	$ \phi \bar{L} E $ & $\ell^+\ell^-$ & $1$ & $g^2 \bar{m}_\chi \lesssim 2 \times 10^{-56}$ \\
& 	$\tilde{\phi} \bar{Q} U$, $\phi \bar{Q} D$ & $q\bar{q}$ & $1$ & $g^2 \bar{m}_\chi \lesssim 6 \times 10^{-57}$ \\
\hline
\multirow{ 4}{*}{$(3)_0$}
& 	$\phi^a \tilde{H}\sigma^a H$&  $hh$, $Z^0 Z^0$,$W^+ W^-$,$f\bar{f}$ & $1:1:2:16N_c y_f^2 \frac{v^2}{m^2_\chi}$ & $\bar{m}_\chi/\bar{\Lambda}^2\gtrsim 9 \times 10^{79}$ \\
&      $\phi^aW^a_{\mu\nu}B^{\mu\nu}$ & $\gamma\gamma$, $Z^0 \gamma$, $Z^0 Z^0$ &  $c_W^2 s_W^2: 2\big(c_W^2-s_W^2\big)^2: c_W^2 s_W^2$   & $\bar{\Lambda}^2/\bar{m}_\chi^3\gtrsim 1\times 10^{31}$  \\
&      $\phi^a \bar{L} E\sigma^a H$ &  $h\ell^+\ell^-$, $Z^0\ell^+\ell^-$, $W^\pm \ell^\mp \nu$, $\ell^+\ell^-$  & $1:1:2:32\pi^2 \frac{v^2}{m^2_\chi}$ &$\bar{\Lambda}^2/\bar{m}_\chi^3\gtrsim 4\times 10^{29}$ \\
&      $\phi^a \bar{Q} U\sigma^a \tilde{H}$, $\phi^a \bar{Q} D\sigma^a H$ &  $h q \bar{q}$, $Z^0q \bar{q}$, $W^\pm q' \bar{q}$, $q\bar{q}$ & $1:1:2:32\pi^2 \frac{v^2}{m^2_\chi}$ & $\bar{\Lambda}^2/\bar{m}_\chi^3\gtrsim 1\times 10^{30}$\\[2pt]
 \hline
\multirow{ 1}{*}{$(3)_1$}
 & 	$\phi^a L^T\sigma^a \sigma^2 L$&  $\nu\nu$ & 1 
 & $g^2 \bar{m}_\chi \lesssim 2 \times 10^{-56}$ \\
\hline\hline
 \multicolumn{5}{|c|}{spin 1/2}\\
 \hline
$(1)_0$ & $\tilde{H} \bar{L} \psi $ & $\nu h$, $\nu Z^0$, $\ell^\pm W^\mp$ & $1:1:2$ & $g^2 \bar{m}_\chi \lesssim 2 \times 10^{-56}$ \\
\hline
 $(2)_{1/2}$ & $\tilde{H} \bar{\psi} E $ &  $\nu h$, $\nu Z^0$, $\ell^\pm W^\mp$ & $1:1:2$ &  $g^2 \bar{m}_\chi \lesssim 2 \times 10^{-56}$ \\
\hline
 $(3)_0$ & $ H \bar{L}\sigma^a \psi^a$ & $\nu h$, $\nu Z^0$, $\ell^\pm W^\mp$ &$1:1:2$ & $g^2 \bar{m}_\chi \lesssim 2 \times 10^{-56}$ \\
\hline\hline
\multicolumn{5}{|c|}{spin 1}\\
 \hline
\multirow{ 2}{*}{$(0)_0$}
& $\bar{f}\gamma_\mu V^{\prime \mu}f$  & $f\bar{f}$ & see text & $N_c g^2 \bar{m}_\chi \lesssim 2 \times 10^{-56}$\\
& $B_{\mu\nu} F^{\prime \mu\nu}/2$    & $f\bar{f}$ & see text & $g^2 \bar{m}_\chi \lesssim 4 \times 10^{-56}$ \\
\hline
\end{tabular}
\caption{
A summary of the different operators that couple a decaying DM candidate to the SM fields. $f$ stands for any of the SM fermions, $q^{(\prime)}$ stands for quarks and $\ell$ for the leptons. We define $\bar{m}_\chi = m_\chi/{\rm PeV}$ and $\bar{\Lambda} = \Lambda/m_{\rm Pl}\,$.}
\label{tab:ModelBuilding}
\end{table}

For context, we begin our discussion of consistent models for DM decay by providing some estimates to correlate the operator description to the DM lifetime.  Specifically, we compute the partial width of a scalar decaying to $n$ different massless scalars, mediated by an $(n+1)$-dimension operator: 
\begin{align}
	\Gamma_{n-{\rm body}} &\sim \frac{m_\chi }{16\pi} \frac{1}{(n-1)!(n-2)!\left(16\pi^2 \right)^{n-2}} \left( \frac{m_\chi }{\Lambda} \right)^{2(n-3)}\, , \\
	\quad \Rightarrow \quad
	\tau_{2-{\rm body}}  &\sim 6 \times 10^{-54}\, {\rm s} \, \left( \frac{m_\chi}{\rm PeV} \right) \left( \frac{m_{\rm Pl}}{\Lambda} \right)^2 \, , \nonumber \\
	\tau_{3-{\rm body}}  &\sim 1 \times 10^{-26}\, {\rm s} \, \left( \frac{\rm PeV}{m_\chi} \right)  \, , \nonumber \\
	\tau_{4-{\rm body}}  &\sim 6 \times 10\, {\rm s} \, \left( \frac{\rm PeV}{m_\chi} \right)^3 \left( \frac{\Lambda}{m_{\rm Pl}} \right)^2 \, , \nonumber \\	
	\tau_{5-{\rm body}}  &\sim 6 \times 10^{29}\, {\rm s} \, \left( \frac{\rm PeV}{m_\chi} \right)^5 \left( \frac{\Lambda}{m_{\rm Pl}} \right)^4 \, ,
\label{eq:taus}
\end{align}
where the phase space integration is taken from~\cite{Kersevan:2004yh}.  In order to incorporate the appropriate mass dimensions for the couplings, we include factors of $\Lambda$, which is expressed in units of $m_\text{Pl}$; an operator carries a factor of $\Lambda^{3-n}$.   

Lifetimes relevant for both IceCube and our gamma-ray constraints are $\mathcal{O}\big(10^{26}-10^{28} \text{ s}\big)$.  Taking the scale where the effective operator is generated to be the Planck scale, along with $m_\chi \sim $ PeV, Eq.~\eqref{eq:taus} shows that this timescale is reproduced if the decay proceeds via a dimension six operator, or if there is additional suppression of a lower dimension interaction due to a small coupling.  The case of a singlet scalar with $\chi \rightarrow \nu\,\nu\, h \,h$ (discussed in the next subsection) is an example of the first type.  The hidden sector glueball is an example of the second, although there the suppression occurs as a consequence of dimensional transmutation---the effective operator connecting the dark sector to the SM is dimension six at scales above $\Lambda_D$, while below the dark confinement scale the operator that connects the glueball to the SM is dimension three with a suppressed coupling. Note a similar scaling was discussed in~\cite{Arvanitaki:2008hq,Arvanitaki:2009yb}.

When constructing a fully consistent theory of the DM interactions, certain decay channels become correlated due to restrictions on the allowed interactions (\emph{e.g.}~from imposing gauge invariance). An expectation for correlated channels can be derived by performing an operator analysis; these results are summarized in Tab.~\ref{tab:ModelBuilding}. For concreteness, we assume that the DM candidate is uncharged, color neutral, and has spin $\le 1$ (a specific spin $3/2$ gravitino model is provided later in this section).  In addition, we impose CP conservation and take all flavor couplings to be diagonal. 

For a given model, we specify all possible renormalizable interactions with the SM fields. We present the lowest dimension operator (up to dimension six) which leads to a final state of interest, such that the branching ratio~(BR) does not vanish as $m_\chi\to \infty$. We comment that in the survey below we have not performed a detailed phenomenological study, so some models could suffer additional particle physics constraints which are not incorporated here.

We use the following notation to specify the quantum numbers of a state:  
\begin{align}
X \sim \big(R_{SU(2)}\big)_Y\,,
\end{align}
where $R_{SU(2)}$ gives the representation under $SU(2)_W$ and $Y$ is the hypercharge.
Then the SM Higgs is denoted as
\begin{align}
	H \sim (2)_{1/2} \, ,
\end{align}
with a vacuum expectation value~(VEV) $v=246\,$GeV.
The SM lepton fields are given by
\begin{align}
	L\sim(2)_{-1/2}\, , \ 
	E\sim(1)_{-1}\, , \	
\end{align}
and the quark fields are specified by
\begin{align}	
	Q\sim(2)_{1/6}\, , \ 
	U\sim(1)_{2/3}\, , \	
	D\sim(1)_{-1/3}\, .		
\end{align}
The $SU(3)_C$ representation is implicit, and $SU(2)_W$-doublets\,(singlets) are left\,(right)-handed fields.  In all cases, flavor indices are suppressed. The field-strength tensors for hypercharge, weak, and strong forces are denoted as  $B_{\mu\nu}$, $W_{\mu\nu}$ and $G_{\mu\nu}$, respectively, and their duals are denoted with a tilde. In addition, we use  $\tilde{H}=i\,\sigma_2H^*$. The different DM candidates are characterized by their electroweak quantum numbers using the same notation. In cases where an additional mass scale is needed (such as a cut-off), it is denoted as $\Lambda$, and we use $g$ for dimensionless extra parameters. In some of the cases the DM candidate is part of larger electroweak multiplet. Therefore, we denoted the full multiplet using $\phi$, $\psi$ or $V$ (depending on the spin), and reserve $\chi$ for the DM candidate itself.  

With the above notation in mind, an extensive EFT classification of decaying DM models is given in Tab.~\ref{tab:ModelBuilding}.  We have organized this table by the spin of the DM, beginning with spin 0, and then further sub-divided into $\left(R_{SU(2)}\right)_Y$ representations.  For each operator, we show the SM final states resulting from the DM decay.  We then give the branching ratios to these states in the limit where the DM mass is much larger than the TeV scale.  Further we provide the requirements on $\bar m_\chi \equiv m_\chi / \text{PeV}$ and $\bar \Lambda = \Lambda / m_\text{Pl}$ in order to have $\tau \gtrsim 10^{27}$ s.  When the operators are dimension $4$, we instead give the requirement in terms of $\bar m_\chi$ and the dimensionless coupling $g$.   

One example that demonstrates why it is important to frame constraints in terms of decay operators derives from model building for the $\nu \nu$ final state relevant for IceCube.  We will argue that it is not possible to write down a PeV mass DM state and corresponding decay operator for which the branching ratio is dominated by the $\nu \nu$ channel, without introducing states with non-trivial quantum numbers (which implies additional new states at similar masses).

First, consider singlet scalar DM $\chi$.  In general, $\chi$ can be coupled to SM gauge neutral operators $\CO_{\rm SM}$ through the operator $\chi \times \CO_{\rm SM}$.\footnote{If $\chi$ is charged under a beyond-the-SM symmetry, it can couple as $\left( \chi^\dagger \chi \right) \times \CO_{\rm SM}$. Then the DM will only decay if $\chi$ acquires a VEV.  This leads to similar phenomenology as in the uncharged case (up to a factor of the $\chi$ VEV divided by the new physics scale).} The different possible operators are summarized in the upper part of Tab.~\ref{tab:ModelBuilding}.   For a single DM state $\chi$, the only operator giving the $\nu\,\nu$ final state is the dimension six operator
\es{sp0}{
\mathcal{L} \supset \frac{1}{\Lambda^2} \chi\, \big(L\,H\big)^2 \,,
} 
which we will take to be restricted to the first generation for simplicity.\footnote{Note that we are assuming dimension five operators of the form $\chi \times \mathcal{L}_\text{SM}$ are suppressed by a symmetry, \emph{e.g.} lepton number.  This requires that the physics at the Planck scale does not badly break this symmetry.}  As shown in Tab.~\ref{tab:ModelBuilding}, at $\sim$PeV masses, the final state $\nu\, \nu\, h\, h$, along with the other channels required to satisfy the Goldstone equivalence theorem, dominate, even though this is a four body decay.  This behavior can be understood by realizing that at these masses it is more appropriate to consider the electroweak symmetry preserving limit of the theory, where this is the only allowed decay mode.   
We will return to this operator in the next subsection, where we show that due to the additional final states is is strongly constrained by {\it Fermi} (see Fig.~\ref{Fig: SM results}).  Finally, we include all other options for singlet scalar decay following our guidelines specified above.  Note that the operators  $\chi\, \psi^\dag D_\mu \sigma^\mu \psi$ and $\chi\, H^\dag D_\mu D^\mu H$ can be mapped onto the operators in Tab.~\ref{tab:ModelBuilding} using the equations of motion.

Next, we investigate the case that the DM candidate is the neutral component of a non-trivial $SU(2)_W$ multiplet, for example a doublet or a triplet. Staring with the doublet case, note that $m_\chi\gg m_h$ is the decoupling limit of two Higgs doublet model, which has been studied extensively---here we follow the conventions and notation of~\cite{Gunion:2002zf}. The three heavy Higgs states, the scalar, pseudo-scalar and charged Higgs, all have similar masses up to electroweak breaking effects set by the scale $v$. There are two valid DM candidates, the scalar and the pseudo-scalar. The spectrum and dominant decay modes depend on the details of the scalar potential as well as the choice of Yukawa couplings between the SM fermions and the two Higgs doublets. For concreteness, we specify three representative scenarios for models where the scalar is the DM: 
(i)~$V_{\hat{\lambda}}$: Three-body decays $\chi\to h\,h\,h, h\,Z^0\,Z^0, h\,W^+\,W^-$;
(ii)~$V_{c_{\beta-\alpha}}$: Two-body decays $\chi\to W^+\,W^-/Z^0\, Z^0$; 
(iii)~Yukawa:  Two-body decays $\chi \to f\,\bar{f}$.
The notation $V_{X}$ is shorthand for the regions of the parameter space where the $X$ coupling dominates the decay phenomenology, following the notation of~\cite{Gunion:2002zf}.  This is summarized in Tab.~\ref{tab:ModelBuilding}.  If the DM is the pseudo-scalar instead, then similar modes are allowed with the appropriate replacements to conserve CP, \emph{e.g.}~$\chi \to Z^0\,h\,h$.     Finally, we emphasize that it is not possible to get direct decays to neutrinos for this model.  

Next, we discuss the case where the DM is the neutral component of an $SU(2)_W$ triplet  $\phi^a$ with hypercharge $Y=1$; for simplicity we assume it does not obtain a VEV.
Given our focus on neutrino final states, the following allowed interaction is particularly interesting
\begin{align}
\mathcal{L} \supset g \, \phi^a L^T \sigma^a \sigma^2 L\,,
\end{align}
where $g$ is a new Yukawa coupling and $\sigma^a$ are the Pauli matrices acting on the $SU(2)_W$ indices.  This operator implies that $\phi^a$ has lepton number two, and yields a 100\% branching fraction for $\chi \rightarrow \nu\, \nu$.  
Since this operators requires the introduction of a full triplet multiplet to preserve gauge invariance, there must also be a charged and doubled charged scalar, with the same mass up to electroweak breaking corrections.  
The possible interaction and decay channels are given in Tab.~\ref{tab:ModelBuilding}.

We conclude our discussion of scalar DM candidates with one more example.  If the neutrinos are Dirac fermions, then the low energy effective theory must include right handed neutrino states $\nu_R$, and one could decay a singlet scalar via the operator $\mathcal{L} \supset \chi\, \nu^T_R\, \nu_R$~\cite{Roland:2014vba}.  Due to neutrino oscillations, these right handed neutrinos will yield a line signature in active neutrinos.  This example is different from $\chi \rightarrow \nu\, \nu$ directly because there are no electroweak corrections to the prompt decay into right handed (sterile) neutrinos, which implies that \emph{Fermi} does not set a relevant constraint.  

We also consider the case where the DM particle has spin.  Fermionic DM can be coupled to the SM via a Yukawa-like interaction. We consider the choices where the DM candidate is an $SU(2)_W$ singlet, doublet, or triplet, with the appropriate hypercharges chosen so that one of the states in the multiplet is a singlet (see Tab.~\ref{tab:ModelBuilding}).  In case of massive spin-1 $Z'$ DM, we assume that it is a singlet under the SM gauge group and acquires all (or most) of its mass from a new scalar VEV.  We consider two possibilities in Tab.~\ref{tab:ModelBuilding}, (i)~one or more of the SM fermions is charged under an additional broken $U(1)$ gauge symmetry; and (ii)~kinetic mixing with the hypercharge gauge boson, see \emph{e.g.}~\cite{Cassel:2009pu,Cline:2014dwa} for some studies of this DM candidate.  It is also possible that this particle is the manifestation of a broken gauging of some subset of the SM global symmetries, \emph{e.g.} the $U(1)$ of $B-L$.  In such a scenario, the resulting gauge coupling must be extremely small to realize $\tau$ in the range of interest.  Finally, for spin-$3/2$ DM, we will consider a specific $R$-parity violating model as discussed in the next sub-section.

\subsection{Additional Models}
\label{sec: add models} 

In this section, we give limits on two additional DM models of interest beyond the example of a hidden sector glueball decaying via the operator $\lambda_D  \,G_{D\mu\nu}\,G_D^{\mu \nu}\, |H|^2 /\Lambda^2$ discussed in the main Letter.

Gravitino DM whose decay is due to the presence of bi-linear $R$-parity violation (via the super-potential coupling $W \supset H_u\, L$) is a well studied scenario. If the gravitino, denoted by $\psi_{3/2}$, is very heavy, it will decay via the following four channels: $\psi_{3/2}~\rightarrow~\nu\,\gamma, \nu \,Z^0, \nu \,h, \ell^\pm \,W^{\mp}$~\cite{Ishiwata:2008cu, Grefe:2008zz}.  For $m_{3/2}$ near the weak scale, the branching ratios are somewhat sensitive to the details of the SUSY breaking masses.  However, once $m_{3/2} \gg v$, the decay pattern quickly asymptotes to $1:1:2$ for the $\nu \,Z^0, \nu\, h, \ell^\pm\, W^{\mp}$ channels respectively, as expected from the Goldstone equivalence theorem.

\begin{figure}[htb]
	\leavevmode
	\begin{center}$
	\begin{array}{cc}
	\scalebox{0.4}{\includegraphics{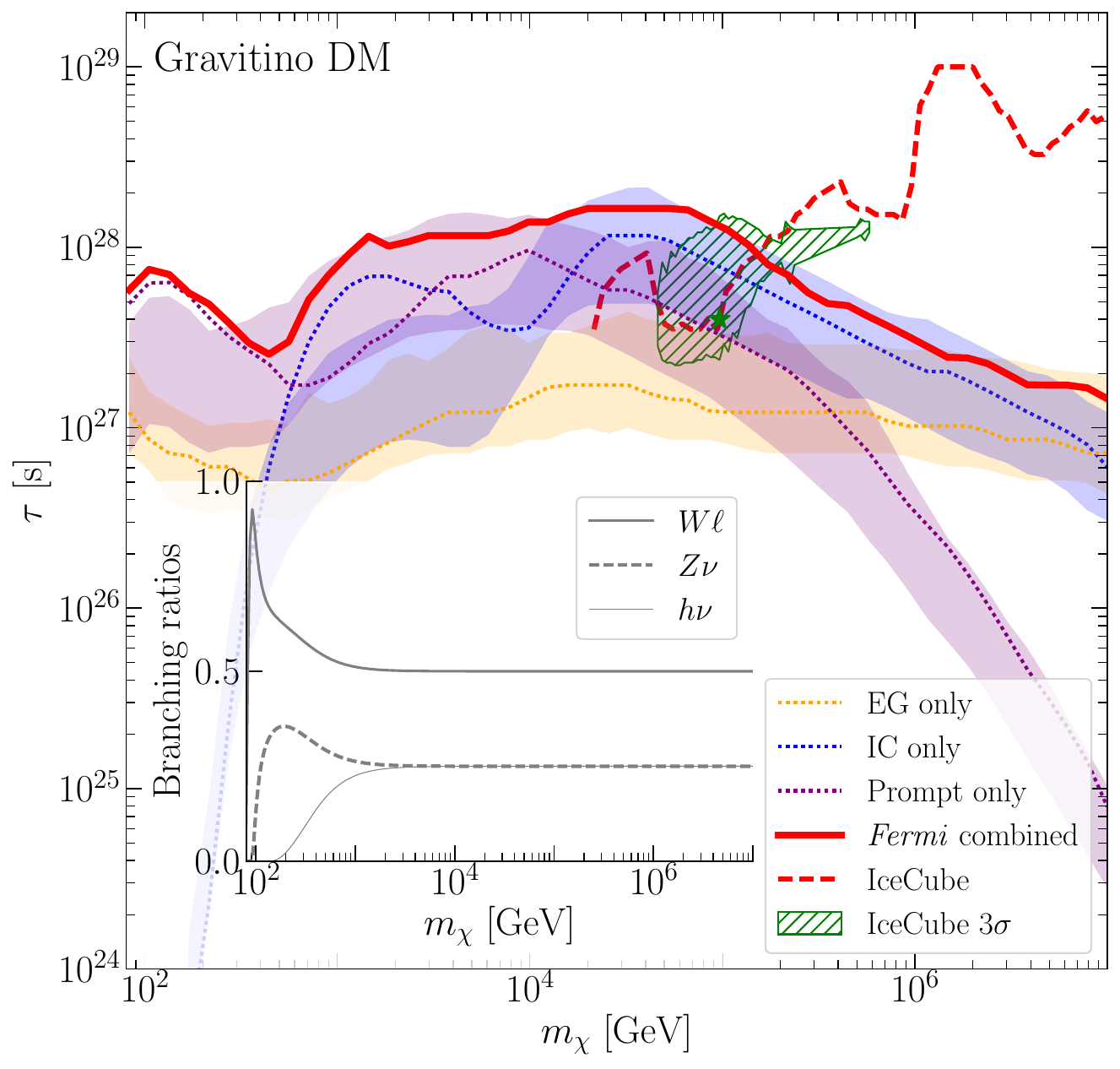}} &\scalebox{0.4}{\includegraphics{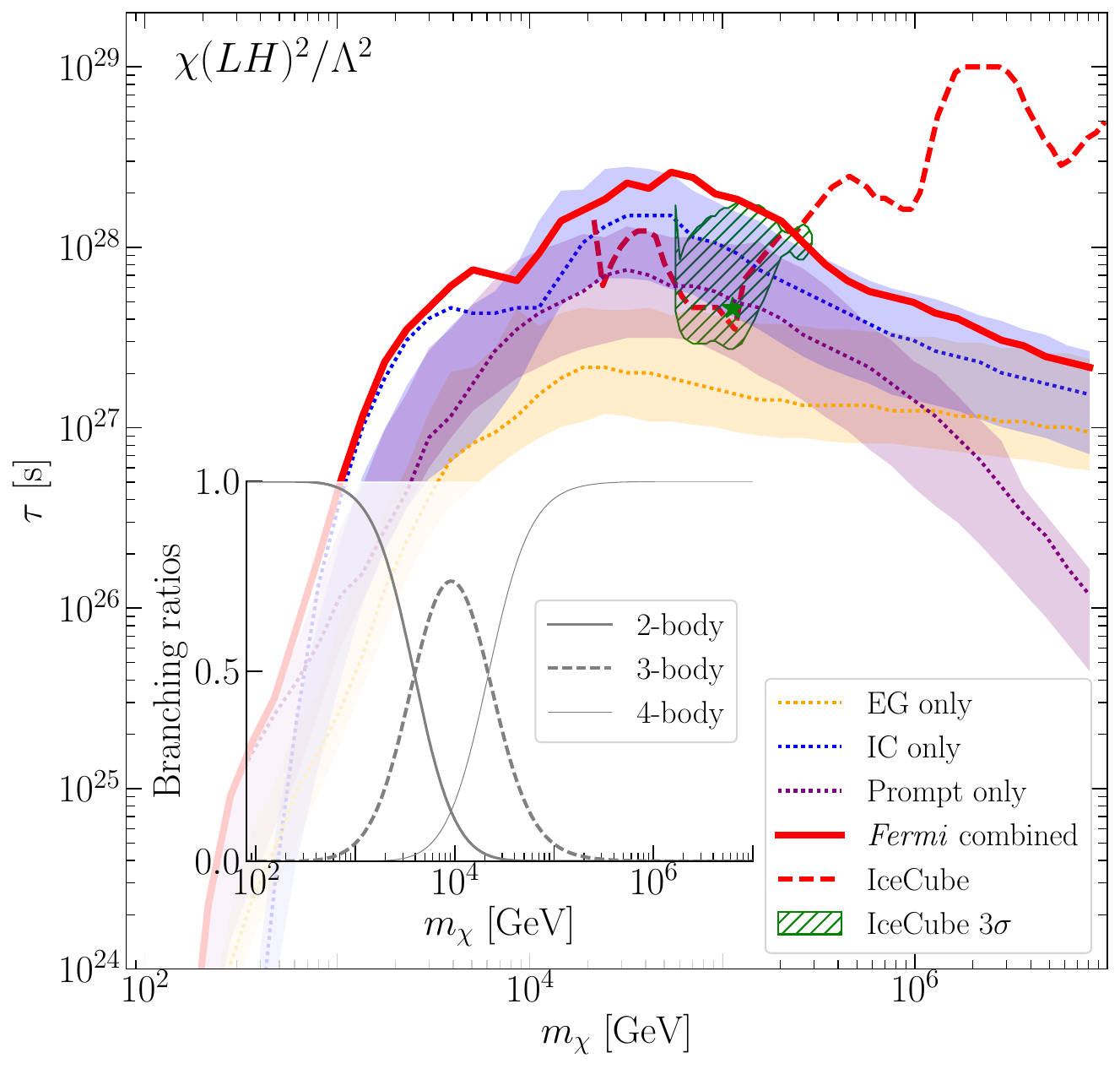}}
	\end{array}$
	\end{center}
	\vspace{-.70cm}
	\caption{Constraints on decaying gravitino DM (left) and DM decaying via the operator $\chi \,(L\,H)^2$ (right).  Notation and labeling is as in Fig.~\ref{Fig: glue} in the main Letter.}
	\vspace{-0.15in}
	\label{Fig: SM results}
\end{figure}

In Fig.~\ref{Fig: SM results} we show the constraints on decaying gravitino DM assuming the above decay modes, with branching ratios given as functions of mass in the inset, using the benchmark parameters of~\cite{Ishiwata:2008cu}.  At masses below the electroweak scale, the $\gamma \nu$ final state dominates.  This channel is best searched for using a gamma-ray line search, which is beyond the scope of this work. For this reason, we only show our constraints for masses above $m_W$.  Note that the region where decaying DM could provide a $\sim$3$\sigma$ improvement over the null hypothesis for the IceCube, the ultra-high-energy neutrino flux (green hashed region) is almost completely excluded by our gamma-ray constraints.  The IceCube constraints, determined using the same methods discussed in the main body of this work, begin to dominate at scales above $\sim$$100$\,TeV.  

In Fig.~\ref{Fig: SM results}, we also show limits obtained on the lifetime of the DM $\chi$ under the assumption that $\chi$ interacts with the SM through the operator in Eq.~\eqref{sp0}---this model was discussed in detail in the previous subsection, see Tab.~\ref{tab:ModelBuilding}. The inset plot shows the branching ratios as a function of energy, and illustrates the transition from two- to three- to four-body decays dominating as the mass is increased.  In this case as well, almost the entire range of parameter space relevant for IceCube is disfavored by our gamma-ray limits.

We use \textsc{FeynRules} 2.0~\cite{Alloul:2013bka} to generate the UFO model files, which are then fed to \textsc{MadGraph5\_aMC{@}NLO}~\cite{Alwall:2011uj, Alwall:2014hca} to compute  the parton-level decay interfaced with \textsc{Pythia} for the showering/hadronization of the decays $\chi \to \nu\, \nu\, h \,h \,, \ \nu\, \nu\, Z^0 \,h \, , \ \nu\, \nu\, Z^0 \,Z^0\, , \, \nu\, e^-\, h\, W^+\, , \ \nu\, e^-\, Z^0\,W^+\, , \ e^-\,e^- \,W^+\, W^+$ and $\chi \to \nu\, \nu \,h \, ,\ \nu\,\nu\,Z^0\, ,\ \nu\,e^- \, W^+ $.

\end{document}